\documentclass[twocolumn,superscriptaddress,numberedappendix,twocolappendix]{emulateapj}
\usepackage{amsmath}
\usepackage{subfigure}
\usepackage[utf8]{inputenc}
\usepackage{amssymb}
\usepackage{graphicx}
\usepackage{verbatim,hyperref}
\usepackage{color}
\usepackage{lipsum}
\usepackage{tikz}
\usepackage[american]{circuitikz}
\usepackage{dsfont}
\allowdisplaybreaks

\renewcommand{\bm}[1]{{\mathbf #1}}

\begin{document}

\title{A practical theorem on using interferometry to measure the global $21$ \lowercase{cm} signal}

\author{
Tejaswi Venumadhav\altaffilmark{1}, 
Tzu-Ching Chang\altaffilmark{2}, 
Olivier Dor\'{e}\altaffilmark{3,4},
Christopher M. Hirata\altaffilmark{5}
}

\affil{\\$^1$ School of Natural Sciences, Institute for Advanced Study, Einstein Drive, Princeton, NJ 08540, USA}
\affil{$^2$ Institute of Astronomy and Astrophysics, Academia Sinica, P.O. Box 23-141, Taipei 10617, Taiwan}
\affil{$^3$ California Institute of Technology, Mail Code 350-17, Pasadena, CA 91125, USA}
\affil{$^4$ Jet Propulsion Laboratory, California Institute of Technology, Pasadena, CA 91109, USA}
\affil{$^5$ Center for Cosmology and Astroparticle Physics, The Ohio State University, 191 West Woodruff Lane, Columbus, OH 43210, USA}


\journalinfo{The Astrophysical Journal, 826:116 (16pp), 2016 August 1}
\submitted{Received 2016 January 5; revised 2016 May 16; accepted 2016 May 18; published 2016 July 26}

\begin{abstract}
The sky-averaged, or global, background of redshifted $21$ cm radiation is expected to be a rich source of information on cosmological reheating and reionizaton. However, measuring the signal is technically challenging: one must extract a small, frequency-dependent signal from under much brighter spectrally smooth foregrounds. Traditional approaches to study the global signal have used single antennas, which require one to calibrate out the frequency-dependent structure in the overall system gain (due to internal reflections, for example) as well as remove the noise bias from auto-correlating a single amplifier output. This has motivated proposals to measure the signal using cross-correlations in interferometric setups, where additional calibration techniques are available. In this paper we focus on the general principles driving the sensitivity of the interferometric setups to the global signal. We prove that this sensitivity is directly related to two characteristics of the setup: the cross-talk between readout channels (i.e. the signal picked up at one antenna when the other one is driven) and the correlated noise due to thermal fluctuations of lossy elements (e.g. absorbers or the ground) radiating into both channels. Thus in an interferometric setup, one cannot suppress cross-talk and correlated thermal noise without reducing sensitivity to the global signal by the same factor -- instead, the challenge is to characterize these effects and their frequency dependence. We illustrate our general theorem by explicit calculations within toy setups consisting of two short dipole antennas in free space and above a perfectly reflecting ground surface, as well as two well-separated identical lossless antennas arranged to achieve zero cross-talk.
\end{abstract}


\maketitle

\section{Introduction}
\label{sec:intro}

One of the future frontiers of observational cosmology is the study of the cosmic dark ages that followed cosmological recombination, the formation of the first luminous objects in the Universe, and the subsequent reionizaton of the Intergalactic Medium (IGM) due to radiation emitted by these sources.

The $21$ cm line of neutral hydrogen promises to be the most powerful probe of the IGM at these redshifts (\onlinecite{1979MNRAS.188..791H,1997ApJ...475..429M}; for a comprehensive list, see \onlinecite{2006PhR...433..181F}). This line corresponds to the transition between the singlet and triplet hyperfine levels of atomic hydrogen in its ground electronic state. The net population of these levels is set by the fraction of neutral hydrogen, while their relative population is a sensitive probe of the thermal state and density of the IGM during this period \citep{1990MNRAS.247..510S,2004ApJ...602....1C}. 

We typically deal with the brightness temperature of this line against the CMB. At a given redshift, this brightness temperature has both uniform and fluctuating components on the sky. Depending on the redshift under consideration, these components contain rich information about cosmology (\onlinecite{2004PhRvL..92u1301L,2004ApJ...608..622Z}; for a complete list, see \onlinecite{2012RPPh...75h6901P}), and the complex astrophysics of the sources that determine the IGM's thermal state and neutral fraction \citep{2004ApJ...610..117W,2005ApJ...626....1B,2006ApJ...637L...1K}. 

The uniform or so-called global signal is very sensitive to the first sources of Lyman-$\alpha$ photons, which drive the spin temperature of neutral hydrogen to the IGM's kinetic temperature. It also probes physical mechanisms that heat the IGM and consequently change its kinetic temperature; these can be the first sources of X-rays \citep{2001ApJ...563....1V,2006ApJ...648L...1C,2010MNRAS.401.2635C,2013ApJ...777..118M,2014Natur.506..197F}, or more exotic mechanisms \citep{2011A&A...528A.149M,2013MNRAS.429.1705V,2014MNRAS.438.2664S,2014A&A...570L...3T,2015arXiv150908408S}. 

Several existing and planned radio experiments attempt to measure the fluctuating component of the $21$ cm signal on the sky at lower redshifts using interferometric techniques \citep{2009AAS...21322605W,2013MNRAS.433..639P,2013A&A...556A...2V,2013MNRAS.429L...5B,2015ApJ...809...61A}. This paper deals with the complementary question of measuring the global $21$ cm signal.

The $21$ cm transition has a rest-frame frequency of $1.4$ GHz; its redshifted frequencies corresponding to the Epoch of Reionization (EoR) and earlier are redward of $\approx 140$ MHz. The brightness temperature of the line when measured against the CMB has a complicated redshift dependence through the dark ages and the EoR, but it is generally expected to be of the order of a few tens of milli-Kelvins (see e.g. \onlinecite{2012RPPh...75h6901P}). Current and future experiments that aim to detect the global signal use the autocorrelation function of the output from well-calibrated receivers in order to study the sky temperature as a function of frequency \citep{2010Natur.468..796B,2012AdSpR..49..433B,2013ITAP...61.2540E,2014ApJ...782L...9V,2015ApJ...799...90B,2015PASA...32....4S,2015ApJ...801..138P}. 

A global signal with such a low amplitude and at the low frequencies of interest is technically complicated to measure for several reasons. The first and most debilitating one is foreground radiation. This is largely due to Galactic synchrotron emission over the frequency range of interest, which has contributions from point sources, unresolved extragalactic sources, bremsstrahlung, dust emission, and radio recombination line radiation \citep{2002ApJ...564..576D,2003MNRAS.346..871O,2008MNRAS.388..247D,2010MNRAS.409.1647J}. Even on the cleanest patches of the sky, these dwarf the cosmological signal by four to five orders of magnitude at frequencies $\nu \lesssim 150$ MHz \citep{1944ApJ...100..279R,1950AuSRA...3...19B,1967MNRAS.136..219B,1970AuJPA..16....1L,2010A&A...522A..67B}. Global signal experiments typically excise frequencies corresponding to known radio recombination lines, and attempt to fit out spectrally smooth components from the measured power as a function of frequency (\onlinecite{1999A&A...345..380S}; for alternative approaches, see \onlinecite{2013PhRvD..87d3002L}).

Measurements at even lower frequencies ($\nu \lesssim 50$ MHz)---those at higher redshifts for the $21$ cm line ($z \gtrsim 27$)---are strongly affected by ``local'' foregrounds due to the earth's ionosphere. Its refraction of background sources mixes spatial and frequency structures in the radio sky \citep{2014MNRAS.437.1056V}, and its dynamic fluctuations add ``flicker'' noise \citep{2014arXiv1409.0513D}. Some preliminary attempts have been made to study this contaminant for global signal experiments at higher frequencies \citep{2015RaSc...50..130R}, but ultimately, the possibility remains that it might preclude ground-based measurements at the lowest frequencies (see, however, \onlinecite{2015ApJ...813...18S}).

The second challenge is calibrating the instrument response (e.g., antenna, receiver, and all stages of processing) as a function of frequency. On the receiver end, this requires an understanding of the pipeline's gain, the noise emitted by amplifiers contained within, and the impedance mismatch at the coupling to the antenna. The former two issues are usually solved for by switching between the sky and reference and calibration noise sources at the ground and some known temperature, respectively \citep{2008ApJ...676....1B,2013ExA....36..319P}. An impedance mismatch at the antenna end results in only a fraction of the sky power coupling into the system; moreover, it leads to additional complications whose details depend on the cables' termination at the amplifiers' input. If these are resistively matched, the matching elements emit Johnson noise that shows up in the output along with reflected waves after a time-delay depending on the cable length (these are the `standing noise waves' described in \onlinecite{1978ITMTT..26...34M}). In the case of open termination at the amplifiers' input, the cable forms a resonant cavity and imprints spectral ripples on a smooth synchrotron spectrum \citep{2012RaSc...47.0K06R}. In addition, the bare antenna temperature differs from the sky temperature due to imperfect ground shielding, local radio frequency interference, and emission and scattering by objects on the horizon, such as trees \citep{2008ApJ...676....1B,2013tra..book.....W}. 

Motivated by these challenges, a few methods have been recently proposed that use multiple-element setups to study the global $21$ cm signal. These methods use cross-correlations between the waveforms at readouts attached to different antennas (which are conventionally used to compute visibilities), which are ostensibly not contaminated by receiver noise bias to the same extent as single antenna setups. The first work in this direction was that of \onlinecite{2014arXiv1406.2585M} (hereafter MSU14), who proposed a so-called zero-spacing interferometer using a partially reflecting sheet as a beamsplitter to divide sky radiation into two components, which are then measured by different antennas. \onlinecite{2015MNRAS.450.2291V} (hereafter VKdB15) proposed and implemented an alternative method wherein they used LOFAR to study the spatial structure in the radio sky induced by the occultation of the global signal by the Moon. A third proposal by \onlinecite{2015ApJ...809...18P} (hereafter PLP15), which was further studied in \onlinecite{2015arXiv150502491S}, is to use a more conventional setup (at least within radio astronomy lore) consisting of an array of antennas above a reflecting ground. 

PLP15 phrase their sensitivity in terms of the shape of the antenna beam on the sky.

MSU14 observe that their setup is only sensitive to the global signal if their beamsplitter is lossy. Moreover, the setups described in MSU14 and PLP15 have a nonzero bias due to the local thermal noise originating in the beamsplitter and/or the imperfect ground and cross-talk between the antennas. These analyses mention these contaminants as sources of systematic noise bias that need further consideration. The observable in VKdB15 is sensitive to the difference in the Moon's and the global signal's temperature; from the perspective of estimating the global temperature, the Moon's temperature is a noise bias (indeed, VKdB15 construct a model for the temperature of the Moon).

In this paper, we provide a framework that simultaneously unifies these methods, generalizes the requirement of a lossy beamsplitter in MSU14, and also throws light on the size of the systematic noise bias. In particular, we obtain the important result that the sensitivity to the sky is {\em directly related} to the size of the systematic noise bias. Hence the latter cannot be ``designed away'' without losing sensitivity to the global signal to the same extent. 

Our results are very general in nature and depend only on the linearity and unitarity of the transformation affected by the setup on incoming signals (unitarity is equivalent to energy conservation after any resistive elements have been appropriately dealt with). The basic idea is to replace the notion of a distant sky (with electromagnetic radiation coming in from past null infinity) by an absorbing sphere of some large radius ${\cal R}$, connected to an ensemble of coaxial cables through which thermal noise is inserted. This fictitious alternative is indistinguishable from the original setup to an observer near the origin. We then use concepts from network theory (energy conservation and reciprocity), as applied to these cables and the cables attached to the antennas on or near Earth, to understand the general properties of signals measured by the observer.

The plan of this paper is as follows: we start with Section \ref{sec:formalism}, wherein we describe a formalism for a setup with an arbitrary number of antennas, and how it transforms incident electric fields due to the sky and local thermal noise. We also relate this to conventional radio astronomy definitions. We then prove our theorem and talk about its implications in Section \ref{sec:proof}. We then illustrate the theorem by explicit calculation in a few toy setups in Section \ref{sec:shortdipole}. We consider a specific limiting case from the PLP15 setup---two identical, lossless antennas at large separation configured to avoid cross-talk---in Section \ref{sec:separated}, where we resolve the apparent discrepancy between our theorem and the traditional formula for an interferometer visibility.\footnote{This section was added at the suggestion of the anonymous referee.} We finish with a discussion of our results in Section \ref{sec:discussion}, and collect some technical details into the Appendices.

\section{Formalism for antenna setup}
\label{sec:formalism}

In this paper, we suppose that each element of the interferometer consists of an antenna that couples electromagnetic waves to a cable. Each cable connects to a receiver, which contains an amplifier that measures the voltage on the cable. There may be additional amplifiers (or other elements, such as mixers and local oscillators) further in the processing chain before the signal is digitized. If so, when we discuss ``the'' amplifier, we mean the first one, because the energy conservation arguments at the heart of this paper do not apply to the outputs of amplifiers or other active power-consuming elements.

We consider a setup with a number of antennas in the presence of incident electromagnetic (EM) radiation that is generated by a distribution of sources in the setup's far field. We decompose the input vector potential, $\bm A(\bm r, t)$, into plane wave modes characterized by a set of frequencies $\nu_m$, directions $\hat{\bm n}_a$ (with a pixel index $a$), and polarizations $\alpha$ with polarization vectors $\bm e_{\alpha}(-\hat{\bm n}_a)$ for radiation traveling in the direction $-\hat{\bm n}_a$. Mathematically,
\begin{align}
  \bm A_{\rm incident}(\bm r, t) & = \sum_{m, \alpha, a} 
  \frac{\Omega_a^{1/2}}{\sqrt{2\pi c\cal T}}\Bigl[
   \psi_{\alpha, \rm in}(\nu_m, \hat{\bm n}_a) \bm e_{\alpha}(-\hat{\bm n}_a) \times \nonumber \\ 
  & \hspace{40pt} e^{-2 \pi \nu_m i (\hat{\bm n}_a \cdot \bm r/c + t)} + {\rm c.c.} \Bigr] \mbox{,} \label{eq:decomposition}
\end{align}
where $c$ is the speed of light, $\psi_{\alpha, \rm in}(\nu_m,\hat{\bm n}_a)$ are frequency components, $\Omega_a$ is the solid angle of pixel $a$, and $\cal T$ is some large duration over which we define Fourier modes. In labeling incoming modes from the sky, we use Latin indices from the beginning of the alphabet ($a,b$) to denote sky pixels, and Greek indices from the beginning of the alphabet ($\alpha,\beta$) for polarization.

We choose the pre-factor in Equation  \eqref{eq:decomposition} such that the autocorrelations of the amplitudes, $\psi_{\alpha, \rm in}(\nu_m,\hat{\bm n}_a)$, equal the energies per mode, a property that we demonstrate in Appendix \ref{sec:discrete} for one choice of discretization of the sky. This is in the Coulomb gauge, in which the electric and magnetic fields are only functions of the vector potential, in the absence of charges. The directions $\hat{\bm n}_a$ range in principle over the whole sky, although some directions may not be visible for a given experimental setup (e.g. below the horizon for a ground-based experiment).

Equation \eqref{eq:decomposition} only includes incoming radiation from the sky (i.e., it omits any radiation from oscillating charges on the antenna or the ground). As such, it is the source contribution, rather than the full EM field in the region of the setup. 

For unpolarized (and possibly anisotropic) thermal radiation from the sky with a temperature $T_{\rm s}(\hat{\bm n})$, the energies per mode are
\begin{align}
~~~ & \!\!\!\!
  \langle \psi^\ast_{\alpha, \rm in}(\nu_n, \hat{\bm n}_a) \psi_{\beta, \rm in}(\nu_m, \hat{\bm n}_b) \rangle \nonumber \\
  & = \frac{h \nu_m}{\exp[{h \nu_m/k_{\rm B}T_{\rm s}(\hat{\bm n}_a)}] - 1} \delta_{\alpha \beta} \delta_{m n} \delta_{a b} \nonumber \\
  & \approx k_{\rm B} T_{\rm s}(\hat{\bm n}_a) \delta_{\alpha \beta} \delta_{m n} \delta_{a b} \mbox{,} \label{eq:rj}
\end{align}
where the delta functions on the right-hand side equal unity when the indices are identical and are zero otherwise. In going from the first to the second line we used the Rayleigh-Jeans approximation for frequencies satisfying $h\nu_m \ll k_{\rm B} T_{\rm s}(\hat{\bm n}_a)$. Note that in this paper, $T_{\rm s}$ always stands for the sky temperature and not the system or spin temperatures.

A system of antennas applies characteristic phase shifts to electric fields that are incident on their surfaces and sums them to produce output signals; they also reflect a part of the incident radiation back into the sky. This reflected radiation is described by outgoing modes whose frequency components $\psi_{\alpha, {\rm out}}(\nu_m, \hat{\bm n}_a)$ are defined in the same manner as those of the incoming ones in Equation  \eqref{eq:decomposition}. From the perspective of the setup, these are radiated away to infinity (in the picture in Appendix \ref{sec:discrete}, this is realized by via an absorbing layer at infinity).
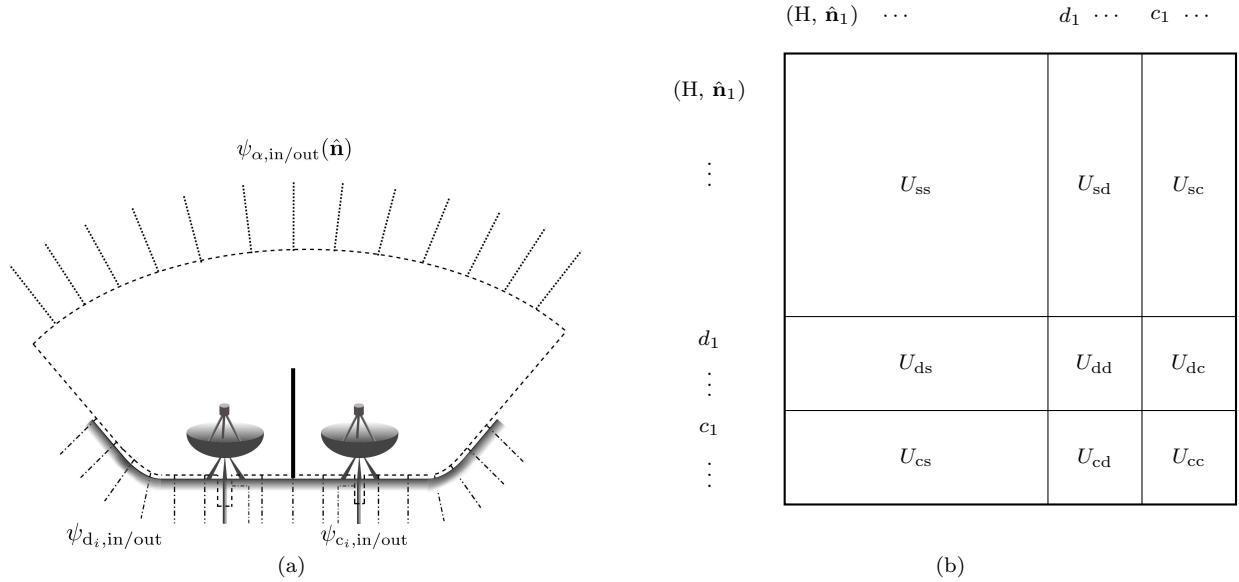
\begin{figure*}[t]
 \centering
  \subfigure[]{
 \includegraphics[width=3in]{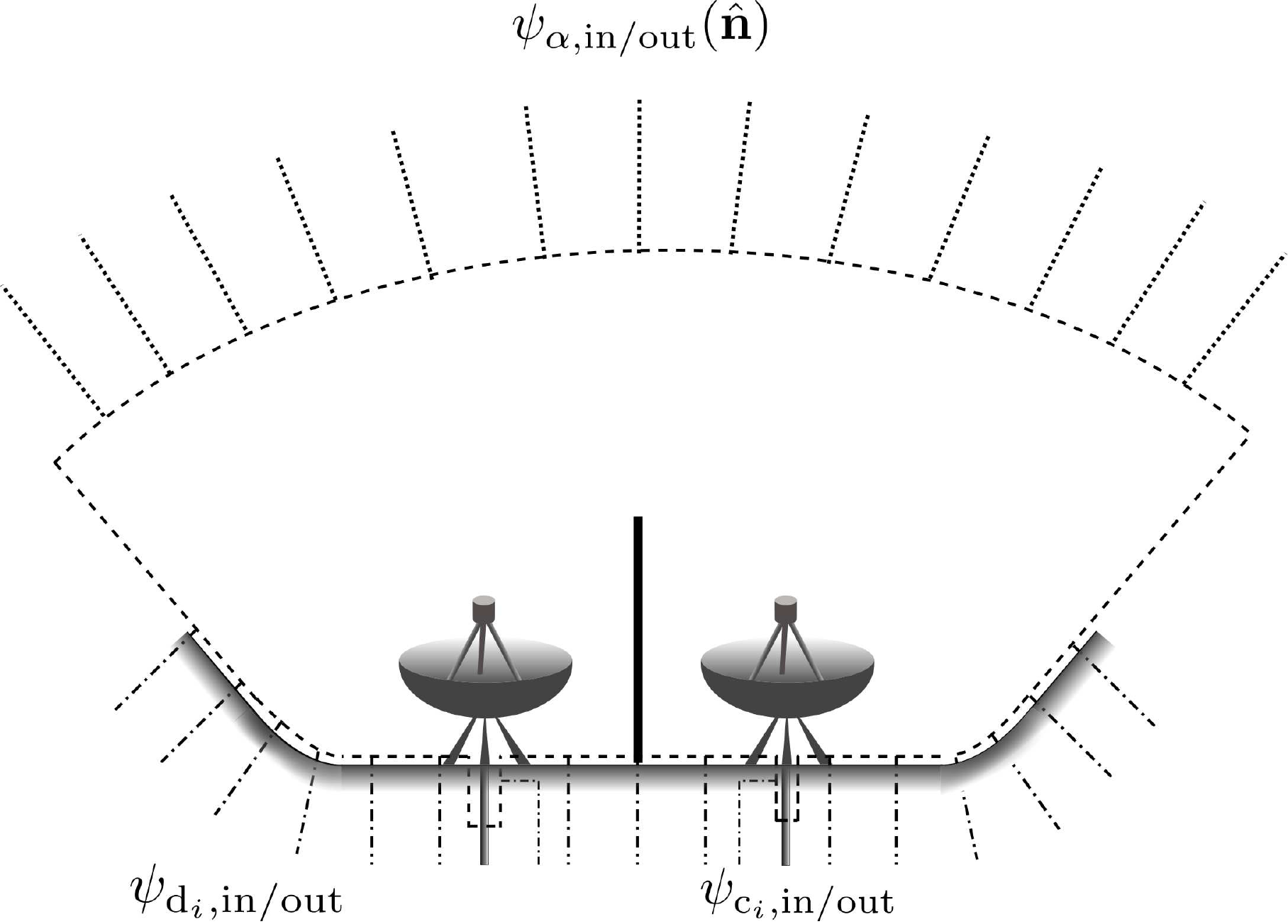}
 \label{fig:schematic}
 }
 \hspace{20pt}
 \subfigure[]{
     \begin{tikzpicture}
     \draw[thick] (-3, -3) rectangle (3, 3);
     \draw (0.5, -3)  -- (0.5, 3);
     \draw (1.75, -3)  -- (1.75, 3);
     \draw (-3, -0.5)  -- (3, -0.5);
     \draw (-3, -1.75)  -- (3, -1.75);     
     \node at (-2.5, 3.5) {(H, $\hat{\bm n}_1)$};
     \node at (-1.5, 3.5) {$\cdots$};
     \node at (0.8, 3.5) {$d_1$};
     \node at (1.3, 3.5) {$\cdots$};
     \node at (2.0, 3.5) {$c_1$};
     \node at (2.5, 3.5) {$\cdots$};
     \node at (-4.0, 2.5) {(H, $\hat{\bm n}_1)$};
     \node at (-4.0, 1.5) {$\vdots$};
     \node at (-4.0, -0.8) {$d_1$};
     \node at (-4.0, -1.3) {$\vdots$};
     \node at (-4.0, -2.0) {$c_1$};
     \node at (-4.0, -2.5) {$\vdots$};
     \node at (-1.25, 1.25) {$U_{\rm ss}$};
     \node at (1.125, 1.25) {$U_{\rm sd}$};
     \node at (2.375, 1.25) {$U_{\rm sc}$};
     \node at (-1.25, -1.125) {$U_{\rm ds}$};
     \node at (1.125, -1.125) {$U_{\rm dd}$};
     \node at (2.375, -1.125) {$U_{\rm dc}$};
     \node at (-1.25, -2.375) {$U_{\rm cs}$};
     \node at (1.125, -2.375) {$U_{\rm cd}$};
     \node at (2.375, -2.375) {$U_{\rm cc}$};
     \node at (0,-3.5) {};
  \end{tikzpicture}
  \label{fig:matrixfig}
 }
\caption{\label{fig:illustrativeschem} Panel (a) shows a schematic depiction of a setup with a number of antennas, such as that in MSU14 or PLP15, as an $n$-port network. It takes inputs from and sends output to a number of ports. The sky ports, s$(\alpha,\hat{\bm n}_a)$, at the top replace celestial sources; their input is the map of sky intensity and their output carries away any radiation emitted from the setup. The cable ports, c$_i$, represent the physical cables connected to each antenna. The dissipation ports, d$_j$, correspond to fictitious cables that replace resistive elements; their input is the thermal (Johnson) noise of each resistor and their output is the power that is dissipated in the resistor in the physical system. Through this replacement, the resulting network conserves energy.
Panel (b) illustrates the scattering matrix for such a network. The $U_{\rm ss}$, $U_{\rm ds}$, and $U_{\rm cs}$ subblocks account for the radiation incident from the sky that is rebroadcasted into the sky, absorbed in dissipative elements, and sent to readout cables, respectively. The $U_{\rm cd}$ subblock accounts for noise due to thermal fluctuations of dissipative elements that enter the readout cables. $U_{\rm cc}$ is the cross-talk matrix whose diagonal entries represent ingoing noise that is reflected back into the same readout cable (such as would occur when the corresponding antenna's feed-point impedance is not matched to the cable), while the off-diagonal entries represent noise broadcasted from one cable+antenna and picked up by the other. If the setup obeys reciprocity, the matrix is symmetric. 
\\
}
\end{figure*}
We assume that outputs from the antennas go to idealized amplifiers (with infinite input impedance) via coaxial cables with impedance $Z_{\rm c}$, which define readout channels ${\rm c}_i$. For simplicity, we assume that the fields in each cable are in the dominant TEM mode, and thus the output voltage signal is
\begin{align}
 ~~~ & \!\!\!
  V_{{\rm c}_i, \rm out}(x,t) \nonumber \\
  & = \sqrt{\frac{Z_{\rm c}}{2\cal T}} \sum_{m} \left[ \psi_{{\rm c}_i, \rm out}(\nu_m) e^{i(\vert \gamma \vert x - 2\pi \nu_m t)} + {\rm c.c.}\right] \mbox{,} \label{eq:cablev}
\end{align}
where the convention for ``in/out" is with respect to the antenna setup (and not the amplifier), and $x$ is a position that is measured along the cable's length and decreases toward the setup. Equation \eqref{eq:cablev} is written for lossless cables with propagation constant $\gamma = i \vert \gamma \vert$ (we will incorporate cable losses later). The pre-factor in Equation  \eqref{eq:cablev} is such that the energy per output mode in the $i$th readout channel is
\begin{align}
  E_{{\rm c}_i, \rm out}(\nu_m) & = \langle \psi_{{\rm c}_i, \rm out}^\ast(\nu_m) \psi_{{\rm c}_i, \rm out}(\nu_m) \rangle \mbox{.} \label{eq:cablep}
\end{align}

Another set of power-sinks are dissipative elements in the setup. These include lossy hardware, cables with finite conductivity, and imperfect ground planes. We model these elements with networks of resistors and purely reactive elements, and replace each resistor by a lossless coaxial cable with an equivalent characteristic impedance that takes energy out of the system. The output signals and energies in the $i$th ``dissipative cable,'' ${\rm d}_i$, are given by Equations \eqref{eq:cablev} and \eqref{eq:cablep} with the appropriate replacements.

Energy is also fed into the setup through incoming modes in both the dissipative cables and readout channels. The former, which we denote by $\psi_{{\rm d}_i, \rm in}(\nu_m)$, are sourced by thermal fluctuations in the electric dipole moments of the dissipative elements according to the fluctuation-dissipation theorem. The latter (i.e., the incoming modes $\psi_{{\rm c}_i, \rm in}(\nu_m)$ at the readout channels) depend on the details of the cables' termination. In this section, we assume that the cables are terminated by purely resistive elements that match the cables' impedance to the idealized amplifiers. In this case, these incoming modes are given by thermal noise, as are the ones in the dissipative cables. We lump the matching elements into the amplifiers and exclude them from the system's description. As we show in Appendix \ref{sec:termination}, the conclusions are unaffected by the choice of termination. 

We assume that all the dissipative elements and terminating resistors radiate into the system at their respective noise temperatures. Mathematically, 
\begin{align}
 ~~~ & \!\!\!\!
  \langle \psi_{({\rm c}/{\rm d})_i, \rm in}^\ast(\nu_m) \psi_{({\rm c}/{\rm d})_j, \rm in}(\nu_m) \rangle \nonumber \\
  & =  E_{({\rm c}/{\rm d})_i, \rm in}(\nu_m) \delta_{ij} = k_{\rm B} T_{({\rm c}/{\rm d})_i} \delta_{ij} \mbox{.} \label{eq:noisetemp}
\end{align}
With the normalizations of the mode functions in Equations \eqref{eq:decomposition} and \eqref{eq:cablev}, the net input/output energy per frequency component is
\begin{align}
  E_{\rm in/out}(\nu_m) 
  & = \sum_{I} \langle \psi^\ast_{I, \rm in/out}(\nu_m) \psi_{I, \rm in/out}(\nu_m) \rangle \mbox{,}\label{eq:energies}
\end{align}
where the capitalized roman index $I$ runs over all the sky modes ($\alpha, \hat{\bm n}_a$), as well as those in the local readout channels ${\rm c}_i$ and dissipative cables ${\rm d}_i$.

Figure \ref{fig:schematic} shows a schematic diagram of the setup, which performs a linear transformation on all its inputs to produce outputs. That is,
\begin{align}
  \psi_{I, {\rm out}} = \sum_{J} U(I; J) \psi_{J, {\rm in}} \mbox{,} \label{eq:scatteringmat}
\end{align}
where the $U(I; J)$ connect the outputs to various source terms, and we have suppressed the frequency $\nu_m$, which is unaffected by linear transformations. In this picture, the setup is an n-port network, and the $U(I; J)$ are elements of its scattering matrix (see e.g. \onlinecite{anen2003radio}). By the reciprocity theorem, the $U(I; J)$ are symmetric in their inputs and outputs (i.e. $U(I; J) = U(J; I)$). In terms of the signals in the sky and the readout/dissipative cables, we have
\begin{subequations}
\label{eq:lineartransform}
\begin{align}
  \psi_{\alpha, \rm out}(\hat{\bm n}_a) & = \sum_{i} U(\alpha,\hat{\bm n}_a;{\rm c}_{i}) \psi_{{\rm c}_{i}, \rm in} + \sum_{i} U(\alpha,\hat{\bm n}_a;{\rm d}_{i}) \psi_{{\rm d}_{i}, \rm in} \nonumber \\
  &~~~ + \sum_{\beta, b} U(\alpha,\hat{\bm n}_a;\beta,\hat{\bm n}_b) \psi_{\beta, \rm in}(\hat{\bm n}_b) \mbox{,} \\
  \psi_{{\rm c}_i, \rm out} & = \sum_j U({\rm c}_i;{\rm c}_j) \psi_{{\rm c}_j, \rm in} + \sum_j U({\rm c}_i;{\rm d}_j) \psi_{{\rm d}_j, \rm in} \nonumber \\
  &~~~ + \sum_{\alpha, a} U({\rm c}_i;\alpha,\hat{\bm n}_a) \psi_{\alpha, \rm in}(\hat{\bm n}_a) \mbox{,} \ {\rm and} \label{eq:beamu} \\
  \psi_{{\rm d}_i, \rm out} & = \sum_j U({\rm d}_i;{\rm c}_j) \psi_{{\rm c}_j, \rm in} + \sum_j U({\rm d}_i;{\rm d}_j) \psi_{{\rm d}_j, \rm in} \nonumber \\
  &~~~ + \sum_{\alpha, a} U({\rm d}_i;\alpha,\hat{\bm n}_a) \psi_{\alpha, \rm in}(\hat{\bm n}) \mbox{.}
\end{align}
\end{subequations}
Figure \ref{fig:matrixfig} shows the various subblocks of the scattering matrix, $U(I; J)$, which describe how ingoing radiation from the sky, dissipative elements, and cables maps to outgoing radiation (or dissipation) in each element.

By construction, the interior of the dashed boundary in Figure \ref{fig:schematic} is free of any dissipation, hence the incoming and outgoing energies are equal (i.e., $E_{\rm in} = E_{\rm out}$ for any input $\psi_{I, {\rm in}}$). If we substitute the relation in Equation \eqref{eq:scatteringmat} for the output signals into the expression for the output energies in Equation \eqref{eq:energies}, and equate the result to the input energies, we get the condition that the $U(I; J)$ form a unitary matrix, i.e.,
\begin{align}
  \sum_{K} U(K; I)^\ast U(K; J) & = \sum_{K} U(I; K)^\ast U(J; K) = \delta_{I J} \mbox{,} \label{eq:unitaryschem}
\end{align}
where the delta on the right-hand side is a Kronecker delta. In terms of the physical modes, some of these conditions are
\begin{subequations}
  \label{eq:unitaryu}
  \begin{align}
    & \sum_{\beta, b} U(\alpha, \hat{\bm n}_a; \beta, \hat{\bm n}_b)^\ast U(\gamma, \hat{\bm n}_c; \beta, \hat{\bm n}_b )\nonumber \\ 
    &~~~ + \sum_{i} U(\alpha, \hat{\bm n}_a; {\rm c}_i)^\ast U(\gamma, \hat{\bm n}_c; {\rm c}_i ) \nonumber \\
    &~~~ + \sum_{i} U(\alpha, \hat{\bm n}_a; {\rm d}_i)^\ast U(\gamma, \hat{\bm n}_c; {\rm d}_i ) = \delta_{\alpha \gamma} \delta_{a c} \mbox{,} \ {\rm and} \label{eq:unitaryusky} \\
    & \sum_{\beta, b} U({\rm c}_i; \beta, \hat{\bm n}_b)^\ast U({\rm c}_j; \beta, \hat{\bm n}_b ) + \sum_{k} U({\rm c}_i; {\rm c}_k)^\ast U({\rm c}_j; {\rm c}_k ) \nonumber \\
    &~~~ + \sum_{k} U({\rm c}_i; {\rm d}_k)^\ast U({\rm c}_j; {\rm d}_k ) = \delta_{i j} \mbox{.} \label{eq:unitaryucable}
  \end{align}
\end{subequations}
We now identify the $U(I; J)$ in terms of more familiar quantities. The quantity $U({\rm c}_i;{\rm c}_j)$ is the cross-talk between the $i$th and $j$th readout channels---if these are connected to different antennas, then this is the signal that is picked up by the $i$th antenna when the $j$th antenna is operated in transmission mode \citep{2002PASP..114...83P}. The quantity $U({\rm c}_i;{\rm d}_j)$ is the part of the thermal noise from the $j$th dissipative element that is picked up by the $i$th antenna. This includes noise generated in all lossy cables, resistive sheets (such as that of MSU14), and by an imperfect ground plane below the setup in PLP15.

The quantity $U({\rm c}_i;\alpha,\hat{\bm n}_a)$ is related to the far-field radiation pattern of the $i$th antenna \citep{1986isra.book.....T}. We now consider a dissipationless single antenna setup with an output cable ${\rm c}_0$ that is connected to a matched load. We substitute Equation \eqref{eq:beamu} in Equation \eqref{eq:cablep}, take $U(\rm c_0; \rm c_0)=0$,\footnote{This is equivalent to the condition that there is no reflected signal when the antenna is used in transmission mode.} and use the mode energies in Equation  \eqref{eq:rj}. The result is
\begin{align}
  E_{{\rm c}_0, \rm out}(\nu_m) & = \sum_{\alpha, a} \vert U(\nu_m, {\rm c}_0; \alpha, \hat{\bm n}_a) \vert^2 k_{\rm B} T_{\rm s}(\hat{\bm n}_a) \mbox{.} \label{eq:ecableout}
\end{align}
The number of pixels ($a_{\rm max}$) and the solid angle per pixel ($\Omega_a$) vary with the discretization procedure. In the limit of small $\Omega_a$, Equation \eqref{eq:ecableout} gives us the usual relation for the received power from an antenna with the effective area to radiation incident from the direction $\hat{\bm n}$ being
\begin{align}
  E_{{\rm c}_0, \rm out}(\nu_m) & = \int {\rm d}\hat{\bm n} \ A(\nu_m, \hat{\bm n}) \frac{\nu_m^2}{c^2} k_{\rm B} T_{\rm s}(\hat{\bm n}_a) \mbox{,} \ {\rm where} \\
  A(\nu_m, \hat{\bm n}) & = \frac{c^2}{\nu_m^2} \lim_{\Omega_a \rightarrow 0} \frac{1}{\Omega_a}  \sum_{\alpha} \vert U(\nu_m, c_0; \alpha, \hat{\bm n}_a = \hat{\bm n}) \vert^2 \mbox{.} \label{eq:effectivea}
\end{align}
Expressed in this language, Equation \eqref{eq:unitaryucable} is equivalent to the usual condition on the effective area (see, e.g., \onlinecite{1966raas.book.....K})
\begin{align}
  \int {\rm d}\hat{\bm n} \ A(\nu_m, \hat{\bm n}) & = \frac{c^2}{\nu_m^2} \mbox{.}
\end{align}
The output from the $i$th antenna in Equation \eqref{eq:beamu} includes two corrections due to the presence of the other antennas: the cross-talk term $U({\rm c}_i;{\rm c}_j)$ can be nonzero, and the far-field radiation pattern $U({\rm c}_i;\alpha,\hat{\bm n})$ (and the effective area $A(\nu_m, \hat{\bm n})$) is distorted due to the presence of the other antennas.

Now we consider a dissipationless two-antenna setup, with cables attached to amplifiers via matched loads (as earlier, we lump the matching resistive load into the amplifier and not into the system). We also take the local noise temperature, $T_{\rm n}$, to zero, so that no noise power is locally input. The measured quantity is the cross-correlation between the signals in the two channels, which we obtain analogously to Equation \eqref{eq:ecableout}:
\begin{align}
~~~ & \!\!\!\!
  \langle \psi_{{\rm c}_1}^\ast (\nu_m) \psi_{{\rm c}_2} (\nu_m) \rangle \nonumber \\
  & = \langle \psi_{{\rm c}_1, {\rm out}}^\ast (\nu_m) \psi_{{\rm c}_2, {\rm out}} (\nu_m) \rangle \nonumber \\
  & = \sum_{\alpha, a} U(\nu_m, {\rm c}_1; \alpha, \hat{\bm n}_a)^\ast U(\nu_m, {\rm c}_2; \alpha, \hat{\bm n}_a) k_{\rm B} T_{\rm s}(\hat{\bm n}_a) \mbox{.}
\end{align}
In general, the transfer-matrix element $U(\nu_m, {\rm c}_j; \alpha, \hat{\bm n}_a)$ has a phase factor $\sim \exp{-(2\pi i \nu_m \bm r_j \cdot \hat{\bm n}_a/c)}$, where $\bm r_i$ is the location of the $i$th antenna.  If we assume that the two antennas are identical, and that their beams are unaffected by the presence of the other, then the other phases cancel out and we get the usual relation for the baseline's visibility with an effective area, as given by Equation \eqref{eq:effectivea}:
\begin{align}
~~~ & \!\!\!\!
  \langle \psi_{{\rm c}_1}^\ast (\nu_m) \psi_{{\rm c}_2} (\nu_m) \rangle \nonumber \\
  & = \int {\rm d}\hat{\bm n} \ A(\nu_m, \hat{\bm n}) \frac{\nu_m^2}{c^2} k_{\rm B} T_{\rm s}(\hat{\bm n}_a) e^{2\pi i \nu_m (\bm r_2 - \bm r_1) \cdot \hat{\bm n}_a} \mbox{.} \label{eq:visibility}
\end{align}

Before we continue, we note the generality of the formalism developed here. The key assumptions used to show that $U$ is unitary were that (i) all elements in the system are linear; (ii) the sources of fluctuations are incident radio waves from the sky, thermal noise from dissipative elements, and any incoming signals in the cables; (iii) the amplifiers on the outgoing signal cables are ideal (in the sense of measuring the true voltage on the cable); and (iv) the problem is time-stationary. To show that $U$ is symmetric, we used the reciprocity theorem, which makes the additional assumption that (v) the system obeys time reversal invariance. Some nonideal behaviors can be easily incorporated into the formalism. For example, any source of noise in the amplifier outputs that is independent of the incoming signal merely adds to the covariance matrix $\langle \psi_{{\rm c}_i}^\ast (\nu_m) \psi_{{\rm c}_j} (\nu_m) \rangle$. An ideal amplifier has infinite input impedance; a finite impedance could be included by modeling the amplifier as a resistor and a reactive element (capacitor or inductor) in parallel with a real amplifier, and setting the effective temperature, $T_{\rm d}$, of that resistor in accordance with the noise power that the amplifier transmits back into the cable.\footnote{The resistance, reactance, and effective temperature would depend on frequency, but our analysis in this paper considers each frequency independently; in particular $U(I;J)$ may depend on frequency.}

Components that break time reversal invariance leave $U$ unitary, but possibly not symmetric (i.e.\ break reciprocity). The most familiar example of such an effect is a material whose electric or magnetic susceptibility is affected by a background (DC) magnetic field (e.g., as used in a Faraday isolator). Because we do not use the symmetry of $U$ in deriving our main theorem (Equation (\ref{eq:theoremstate})), this remains valid even in the presence of such devices. However, the examples that we give in Sections~\ref{sec:shortdipole} and \ref{sec:separated} use reciprocity and would need to be revisited if time reversal-violating components are used.

On the other hand, any sources of nonlinearity (whether in the amplifier or in an upstream component; e.g., a nonlinear material used in the antenna) or signal-dependent noise (e.g., an amplifier whose noise power spectrum increases when the signal is increased) cannot be treated within the matrix formalism described here. While we are unaware of any practical proposals that exploit these effects to measure the monopole sky signal, our theorem would not place any restrictions on such a device.

\section{Response of an interferometer to the monopole of the sky}
\label{sec:proof}

We are now interested in how an interferometer responds to the sky monopole. In what follows, we separate out the monopole by writing
\begin{equation}
T_{\rm s}(\hat{\bm n}_a) = \bar T_{\rm s} + \Delta T_{\rm s}(\hat{\bm n}_a),
\end{equation}
where the sky average of $\Delta T_{\rm s}$ is zero. We also relax the assumptions involved in deriving Equation \eqref{eq:visibility}, i.e., we include dissipative cables $d_{\rm i}$ and assume all local sources have nonzero noise temperatures.

Consider two distinct readout channels in the setup: ${\rm c}_{i}$ and ${\rm c}_{j}$ with $i \neq j$. In a typical interferometric setup, such as the minimal one illustrated in Figure \ref{fig:schematic}, these are cables connecting to different antennas (this can also describe a scenario with multiple readout cables attached to a single antenna). 

The cross-correlation between the waveforms measured in the two readouts is
\begin{align}
  \langle \psi_{{\rm c}_i}^\ast \psi_{{\rm c}_j} \rangle
  & = \langle (\psi_{{\rm c}_i, {\rm in}}^\ast + \psi_{{\rm c}_i, {\rm out}}^\ast )( \psi_{{\rm c}_j, {\rm in}} + \psi_{{\rm c}_j, {\rm out}} ) \rangle \mbox{.}  \label{eq:netcorrelation} 
\end{align}
We now use Equation \eqref{eq:beamu} for the output in each readout channel, and use Equations \eqref{eq:noisetemp} and \eqref{eq:rj} for the input noise at the cable terminations and dissipative cables, and the input energies from the sky to obtain
\begin{align}
  \langle \psi_{{\rm c}_i}^\ast \psi_{{\rm c}_j} \rangle & = U({\rm c}_j; {\rm c}_i) k_{\rm B} T_{{\rm c}_i} + U({\rm c}_i; {\rm c}_j)^\ast k_{\rm B} T_{{\rm c}_j} + \nonumber \\
  & ~~~ \sum_{I} U({\rm c}_i; I)^\ast U({\rm c}_j; I) k_{\rm B} T_{I} \mbox{.}
\end{align}
We use the unitarity constraint of Equation \eqref{eq:unitaryucable} to rewrite this, subtracting 0 times $\bar T_{\rm s}$:
\begin{align}
~~~ & \!\!\!\!
  \langle \psi_{{\rm c}_i}^\ast \psi_{{\rm c}_j} \rangle \nonumber \\
  & = U({\rm c}_j; {\rm c}_i) k_{\rm B} T_{{\rm c}_i} + U({\rm c}_i; {\rm c}_j)^\ast k_{\rm B} T_{{\rm c}_j}  \nonumber \\
  & ~~~ +\sum_{I} U({\rm c}_i; I)^\ast U({\rm c}_j; I) k_{\rm B} (T_{I} - \bar T_{\rm s}) \nonumber \\
  & = U({\rm c}_j; {\rm c}_i) k_{\rm B} T_{{\rm c}_i} + U({\rm c}_i; {\rm c}_j)^\ast k_{\rm B} T_{{\rm c}_j}  \nonumber \\
  & ~~~ +\sum_{k} U({\rm c}_i; {\rm c}_k)^\ast U({\rm c}_j; {\rm c}_k) k_{\rm B} (T_{{\rm c}_k} - \bar T_{\rm s}) \nonumber \\
  & ~~~ +\sum_{k} U({\rm c}_i; {\rm d}_k)^\ast U({\rm c}_j; {\rm d}_k) k_{\rm B} (T_{{\rm d}_k} - \bar T_{\rm s}) \nonumber \\
  & ~~~ +\sum_{\alpha a} U({\rm c}_i; \alpha, \hat{\bm n}_a)^\ast U({\rm c}_j; \alpha, \hat{\bm n}_a) k_{\rm B} \Delta T_{\rm s}(\hat{\bm n}_a)
  \mbox{.} \label{eq:theoremint}
\end{align}
We now take the partial derivative of this expression with respect to $\bar T_{\rm s}$ at fixed instrument properties and a fixed anisotropy map $\Delta T_{\rm s}(\hat{\bm n})$. This leads to the main theorem of this paper
\begin{align}
~~~ & \!\!\!\!
  \frac{1}{k_{\rm B}} \frac{\partial}{\partial \bar T_{\rm s}} \langle \psi_{{\rm c}_i}^\ast \psi_{{\rm c}_j} \rangle \nonumber \\
  & = - \left[ U({\rm c}_i; {\rm c}_i)^\ast U({\rm c}_j; {\rm c}_i) + U({\rm c}_i; {\rm c}_j)^\ast U({\rm c}_j; {\rm c}_j) \right]  \nonumber \\
  & ~~~ -\sum_{k \neq i, j} U({\rm c}_i; {\rm c}_k)^\ast U({\rm c}_j; {\rm c}_k) - \sum_{{\rm d}_k} \frac{1}{k_{\rm B}} \frac{\partial}{\partial T_{{\rm d}_k}} \langle \psi_{{\rm c}_i}^\ast \psi_{{\rm c}_j} \rangle \mbox{.} \label{eq:theoremstate}
\end{align}
The left-hand side is the sensitivity of the cross-correlation of the waveforms at the readout channels to the sky monopole (at fixed anisotropy; i.e., at fixed dipole, quadrupole, etc.). The quantity $\langle \psi_{{\rm c}_i}^\ast \psi_{{\rm c}_j} \rangle$ is the geometric mean of the energy per mode in both the channels, multiplied by the complex correlation coefficient. As we noted in Equation \eqref{eq:visibility}, this is usually used to compute the interferometric visibility.

The right-hand side of Equation \eqref{eq:theoremstate} is the sum of three sets of terms, at least one of which must be nonzero for this visibility to be sensitive to the globally averaged sky temperature, $\bar T_{\rm s}$. In order, these terms require:
\begin{enumerate}
  \item Nonzero cross-talk between the two readout channels (i.e., $U({\rm c}_i; {\rm c}_j) \neq 0$), and that at least one of the antennas is not impedance matched with its cable (i.e., $U({\rm c}_{i/j}; {\rm c}_{i/j}) \neq 0$). 
  \item Nonzero cross-talk between both the channels and at least another readout channel (i.e., $U({\rm c}_i; {\rm c}_k), U({\rm c}_j; {\rm c}_k) \neq 0$, with $k \ne i, j$).
  \item The presence of dissipative elements that can emit noise into both the cables; as a result the cross-correlation picks up a bias (i.e., $(\partial/\partial T_{{\rm d}_k}) \langle \psi_{{\rm c}_i}^\ast \psi_{{\rm c}_j} \rangle \neq 0$ for some $d_{\rm k}$).
\end{enumerate}
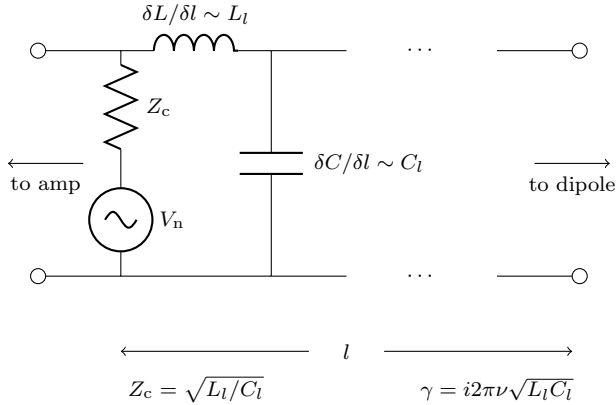
\begin{figure}[t]
  \centering
    \begin{tikzpicture}
      \draw (0,1.5) -- (1,1.5) to [R=$Z_{\rm c}$] (1,0) to [sV=$V_{\rm n}$] (1,-1.5) -- (0,-1.5);
      \draw (-0.1,1.5) circle (0.1);
      \draw (-0.1,-1.5) circle (0.1);
      \draw[->] (0.5,0) -- (-0.5,0);
      \node at (0,-0.3) {to amp};
      \draw (1,1.5) to [L=$\delta L/\delta l \sim L_l$] (3,1.5) to [C=$\delta C/\delta l \sim C_l$] (3,-1.5) -- (1,-1.5);
      \draw (3,1.5) -- (4,1.5);
      \draw (3,-1.5) -- (4,-1.5);
      \node at (5,1.5) {$\cdots$};
      \node at (5,-1.5) {$\cdots$}; 
      \draw (6,1.5) -- (7,1.5);
      \draw (6,-1.5) -- (7,-1.5);
      \draw (7.1,1.5) circle  (0.1);
      \draw[->] (6.5,0) -- (7.5,0);      
      \node at (7,-0.3) {to dipole};
      \draw (7.1,-1.5) circle  (0.1);
      \draw[<-] (1,-2.5) -- (3.5,-2.5);
      \draw[->] (4.5,-2.5) -- (7,-2.5);
      \node at (4,-2.5) {$l$};
      \node at (2,-3) {$Z_{\rm c} = \sqrt{L_l/C_l}$};
      \node at (6,-3) {$\gamma = i 2\pi \nu \sqrt{L_l C_l}$};
    \end{tikzpicture}
    \caption{\label{fig:drivecircuit} Equivalent circuit for an idealized cable+receiver. The amplifier is assumed to have an infinite input impedance. The resistive load matches the cable's impedance $Z_{\rm c}$; the impedance and the propagation constant, $\gamma$, are functions of the cable's inductance and capacitance per unit length.}
\end{figure}
MSU14 noted the third condition in a restricted context (see their Sec. III). However, they did not make the connection between the cross-correlation's size and that of the correlated input noise due to the emission originating from their dissipative sheet. In the derivation of Equation \eqref{eq:theoremstate}, we have not explicitly conditioned on the location of the dissipative elements within the setup. Hence, we can apply it to the method of VKdB15 by modeling the Moon as such a dissipative element in the far field of the antennas used to compute visibilities, but within the setup's definition. We see from the last term in Equation \eqref{eq:theoremint} that the cross-correlation of the readouts is naturally sensitive to the difference of the global signal's and the Moon's temperature. The latter is a systematic noise bias from the perspective of measuring the former. PLP15 note that cross-talk between their antennas is a source of systematic noise bias, but assume that it can be mitigated by appropriate design choices or physical separation of the antennas. Equation \eqref{eq:theoremstate} shows that any such steps will reduce the sensitivity to the signal by the same factor. This will become clearer in the example that follows.

In the examples of our theorem that follow, we focus on the case of a uniform sky and therefore make the replacement $\bar T_{\rm s}\rightarrow T_{\rm s}$, because the anisotropy map $\Delta T_{\rm s}(\hat{\bm n})$ does not appear in our main theorem (Equation \ref{eq:theoremstate}). However, one should remember that Equation (\ref{eq:theoremstate}) remains valid even when the sky temperature is anisotropic. The role of anisotropies is made clear by Equation (\ref{eq:theoremint}): the observed correlation is the sum of that which would be observed for a uniform sky at the monopole temperature $\bar T_{\rm s}$, plus a term associated with the anisotropy map that does not depend on $\bar T_{\rm s}$.

\section{Setups with two short dipole antennas}
\label{sec:shortdipole}

In this section, we work out the examples of interferometers with two parallel, side by side short dipole antennas in free space, and above a perfectly reflecting ground. These are not practical setups, but rather ones within which we can illustrate our theorem by explicit calculation. 

\subsection{Dipoles in free space}
\label{subsec:freespace}

We first start with the free-space case. The assumption of short dipoles helps us in two ways. First, short dipoles have a small radiation resistance ($\sim O(kd)^2 Z_0$, where $k = 2\pi \nu/c$ is the wavenumber, $d$ is the dipoles' size, and $Z_0$ is the impedance of free space). The effect of radiation is a small perturbation to the electric fields in short dipoles' vicinity; hence, their response to incident fields is essentially electrostatic in nature. Secondly, we can completely describe their fields using a single parameter; their dipole moment. 

From the perspective of the receiving circuit, the short dipoles' behavior is dominantly capacitive; we assume they have a capacitance $C$.  For a given stored charge $Q$, they develop dipole moments ${\bm p} = \xi Q \hat{\bm x}$, where $\xi$ is a conversion factor with units of length ($\xi \ll \lambda = c/\nu$) and we have oriented the dipole along the $x$ direction. Their radiative behavior is a perturbation in terms of the small parameter $k \xi$.
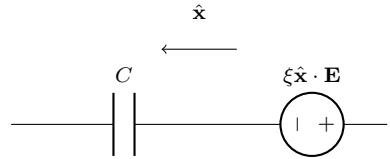
\begin{figure}[t]
 \centering
  \begin{tikzpicture}
    \node at (1.5,3.5) {$\hat{\bm x}$};
    \draw[<-] (1,3) -- (2,3);
    \draw (-1,2) to [C=$C$] (2,2) to [V=$\xi \hat{\bm x} \cdot {\bm E}$] (4,2);
  \end{tikzpicture}
 \caption{\label{fig:eqcircuit} Th\'evenin equivalent for a short dipole placed in an external electric field, ${\bm E}$. The dipole has a capacitance $C$ and a conversion factor $\xi$.}
\end{figure}

We assume that the dipoles are coupled to coaxial cables, ${\rm c}_1$ and ${\rm c}_2$, via baluns. As earlier, the cables are lossless transmission lines with propagation constant $\gamma = i \vert \gamma \vert$, terminated at idealized amplifiers by matching resistive loads. We assume that these loads are at some noise temperature $T_{\rm n}$. The cables have a characteristic impedance, $Z_{\rm c}$, and length, $l$. Figure \ref{fig:drivecircuit} shows the receiving circuit.

If we measure the noise voltage $V_{\rm n}$ in Figure \ref{fig:drivecircuit} over a time-interval ${\cal T}$ and define Fourier modes with frequency $\nu_m = m/{\cal T}$, we have
\begin{equation}
  V_{\rm n}  = \sqrt{\frac{Z_{\rm c}}{2\cal T}} \sum_m {\cal V}_{\rm n}(\nu_m) e^{-2\pi i \nu_m t} + {\rm c.c.} \mbox{,}
    \label{eq:noisefft}
\end{equation}
with
\begin{equation}
 \langle {\cal V}_{\rm n}^\ast(\nu_m) {\cal V}_{\rm n}(\nu_m) \rangle  = 4 k_{\rm B} T_{\rm n} \mbox{.}
\end{equation} 
We choose the factor of $1/\sqrt{2}$ to obtain the right normalization for the frequency components (see discussion around Equation \eqref{eq:cablemodes}). The {\em input} noise signal (i.e. before considering any reflections at the dipole end) has a voltage drop that is half of this noise voltage because the load is matched to the line. This gives us the right normalization for the input modes and noise energies with a common noise temperature $T_{\rm n}$ (see Equations \ref{eq:cablev} and \ref{eq:noisetemp}).

By considering the reflection of propagating modes at the dipole, and the resulting relation between the incoming and outgoing amplitudes at the resistor, we have
\begin{align}
  U({\rm c}_i; {\rm c}_i) & = e^{2 i (\vert \gamma \vert l + \phi_{\rm C})} + O(k \xi)^2 \mbox{,} \ {\rm where}\label{eq:selfrefl}\\
  \phi_{\rm C} & = \arctan{(2 \pi \nu C Z_{\rm c})} \mbox{.} \label{eq:reflectps}
\end{align}
The small correction of $O(k\xi)^2$ in Equation \eqref{eq:selfrefl} is due to the dipole's radiation, which results in a broadcasted electric field ${\bm E}({\bm r}, t; {\rm c}_i)$. Measured over a time-interval $\cal T$, we define the Fourier modes of this field by
\begin{align}
  {\bm E}({\bm r}, t; {\rm c}_i) & = \sqrt{\frac{1}{2\mathcal{T}}} \sum_m {\bm E}({\bm r}, \nu_m; {\rm c}_i) e^{-2\pi i \nu_m t} + {\rm c.c.} \label{eq:efourier}
\end{align}
For a given input $\psi_{{\rm c}_i, {\rm in}}$ we have a dipole moment ${\bm p}_i(\nu_m)$, given by
\begin{align}
  {\bm p}_i(\nu) & = \xi C \sqrt{Z_{\rm c}} e^{i \vert \gamma \vert l} (1 + e^{2 i \phi_{\rm C}}) \psi_{{\rm c}_i, {\rm in}}(\nu) \hat{\bm x} \notag \\
  & = \frac{\xi \sin{(\phi_{\rm C}})}{\pi\nu \sqrt{Z_{\rm c}}} e^{i (\vert \gamma \vert l + \phi_{\rm C})} \psi_{{\rm c}_i, {\rm in}}(\nu) \hat{\bm x} \mbox{,} \label{eq:dipoletransf}
\end{align}
where we have used Equation \eqref{eq:reflectps} to express the capacitance in terms of the angle $\phi_{\rm C}$. Note the factor of $\sqrt{Z_{\rm c}}$ in the first line, which converts voltage back into physical units.

The broadcasted electric fields are dipole fields, given by
\begin{align}
~~~ & \!\!\!\!
  {\bm E}({\bm r}, \nu; {\rm c}_i) \notag \\
  \notag \\[0pt]
  & = \frac{(k r)^2 [(\hat{\bm r} \times {\bm p}_i) \times \hat{\bm r}] + [3 \hat{\bm r} (\hat{\bm r} \cdot {\bm p}_i) - {\bm p}_i](1 - i k r)}{r^3} e^{i k r} \mbox{,} \label{eq:dipoleE}
\end{align}
where we define the displacement vector $\bm r$ with reference to the dipole's location. If the displacement is orthogonal to the dipole moment, we have
\begin{align}
  {\bm E}({\bm r} \perp {\bm p}_i, \nu; {\rm c}_i)
  & = {\bm p}_i \frac{(k r)^2 + i kr - 1}{r^3} e^{i k r} \mbox{.} \label{eq:dipoleEperp}
\end{align}
We also need the dipoles' behavior under the receiving condition. By the reciprocity theorem, the parameter $\xi$ governs both the transmission and receiving properties of the short dipoles. The result is that an ambient electric field effectively adds an extra voltage source in series with the capacitor, with voltage $\xi \hat{\bm x} \cdot {\bm E}$ (we present a more detailed derivation of this in Appendix \ref{sec:reciprocity}). Figure \ref{fig:eqcircuit} shows the Th\'evenin equivalent circuit for a short dipole. 

In units where the square of the waveform is the energy per mode, the outgoing signal at the readout of ${\rm c}_i$ due to an time-varying incident electric field ${\bm E_i}(\nu)$ and reflected noise is 
\begin{align}
  \psi_{{\rm c}_i, {\rm out}}(\nu) & = -\frac{\xi \hat{\bm x} \cdot {\bm E_i}(\nu) \sqrt{Z_{\rm c}}}{(Z_{\rm c} + i/(2 \pi \nu C))} e^{i \vert \gamma \vert l} + U({\rm c}_i; {\rm c}_i) \psi_{{\rm c}_i, {\rm in}}(\nu) \nonumber \\
  & = i \frac{\xi \hat{\bm x} \cdot {\bm E_i}(\nu)}{\sqrt{Z_{\rm c}}}  \sin{(\phi_{\rm C})} e^{i (\vert \gamma \vert l + \phi_{\rm C})} \notag \\
  & ~~~ + \left[ e^{2 i (\vert \gamma \vert l + \phi_{\rm C})} + O(k \xi)^2 \right] \psi_{{\rm c}_i, {\rm in}}(\nu) \label{eq:outunsimpl} \mbox{.}
\end{align}
The incident electric field at each dipole's location is the superposition of the field due to the sky and that due to the other dipole; that is, 
\begin{align}
  {\bm E}_{i}(\nu) & = {\bm E}_{i, {\rm sky}}(\nu) + {\bm E}({\bm r}_i, \nu; {\rm c}_{j}) \mbox{.}
\end{align}
The electric field due to the second dipole is a combination of the reflected sky signal and the broadcasted noise. The first contribution is down by a factor of $(k\xi)^2$, because the second dipole has to absorb and reradiate. We use Equation \eqref{eq:dipoleEperp} for the second contribution with ${\bm r} = {\bm r}_{i j} = {\bm r}_i - {\bm r}_j$. 

Substituting into Equation \eqref{eq:outunsimpl} yields
\begin{align}
  \psi_{{\rm c}_i, {\rm out}}(\nu) & = i \frac{\xi \hat{\bm x} \cdot [ {\bm E}_{i, {\rm sky}}(\nu) + {\bm E}({\bm r}_i, \nu; {\rm c}_{j})]}{\sqrt{Z_{\rm c}}} \nonumber \\
  &~~~ \times \sin{(\phi_{\rm C})} e^{i (\vert \gamma \vert l + \phi_{\rm C})} \nonumber \\
  &~~~ + \left[ e^{2 i (\vert \gamma \vert l + \phi_{\rm C})} + O(k \xi)^2 \right] \psi_{{\rm c}_i, {\rm in}}(\nu) \nonumber \\ 
  & = i \frac{\xi \hat{\bm x} \cdot {\bm E}_{i, {\rm sky}}(\nu)}{\sqrt{Z_{\rm c}}}  \sin{(\phi_{\rm C})} e^{i (\vert \gamma \vert l + \phi_{\rm C})} \nonumber \\
  &~~~ + \left[ e^{2 i (\vert \gamma \vert l + \phi_{\rm C})} + O(k \xi)^2 \right] \psi_{{\rm c}_i, {\rm in}}(\nu) \notag \\
  &~~~ + i \frac{\xi}{\sqrt{Z_{\rm c}}} \left[ \frac{\xi \sin{(\phi_{\rm C}})}{\pi \nu \sqrt{Z_{\rm c}}} e^{i (\vert \gamma \vert l + \phi_{\rm C})} \right. \nonumber \\
  &~~~ \left. \frac{ (k r_{ij})^2 + i k r_{ij} - 1 }{r_{ij}^3} \right. \nonumber \\
  &~~~ \left. \times e^{i k r_{ij}} \psi_{{\rm c}_j, {\rm in}}(\nu) + O(k \xi)^2 \right] \sin{(\phi_{\rm C})} e^{i (\vert \gamma \vert l + \phi_{\rm C})} \mbox{.} \label{eq:psioutsimpl}
\end{align}
The first term is the signal picked up from the sky, the second term is the reflected input thermal noise, and the third term is the cross-talk coefficient, which is the noise broadcasted by the second dipole and picked up by the first. The lowest-order expression for the associated coefficient is
\begin{align}
  U({\rm c}_i; {\rm c}_j) & = i \frac{\xi^2}{\pi \nu Z_{\rm c}} \sin^2{(\phi_{\rm C}}) e^{2 i (\vert \gamma \vert l + \phi_{\rm C})} \times \nonumber \\
  & ~~~ \frac{ (k r_{ij})^2 + i k r_{ij} - 1 }{r_{ij}^3} e^{i k r_{ij}} \mbox{.} \label{eq:crossrefl}
\end{align}
The cross-correlation between the signals at the two short dipoles' terminals is given by Equation \eqref{eq:netcorrelation}. We obtain the sky contribution from the first term in Equation \eqref{eq:psioutsimpl}. 
\begin{align}
~~~ & \!\!\!\!
  \langle \psi^\ast_{{\rm c}_1}(\nu) \psi_{{\rm c}_2}(\nu) \rangle \vert_{\rm sky} \notag \\
  & = \frac{\xi^2}{Z_{\rm c}} \sin^2{(\phi_{\rm C})} \langle [\hat{\bm x} \cdot {\bm E}^\ast_{1, {\rm sky}}(\nu)] [\hat{\bm x} \cdot {\bm E}_{2, {\rm sky}}(\nu)] \rangle \mbox{.} \label{eq:skycontrib1}
\end{align}
We obtain the frequency components of the electric field from the sky using the continuous-sky limit of Equation \eqref{eq:temp2}, while keeping in mind the definition in Equation \eqref{eq:efourier}
\begin{align}
 ~~~ & \!\!\!\!
  {\bm E}_{\rm sky}({\bm r}, \nu) \notag \\
  & = i \sqrt{\frac{4\pi \nu^2}{c^3}} \sum_{\alpha} \int {\rm d}\hat{\bm n} \ \psi_{\alpha, \rm in}(\nu, \hat{\bm n}) \bm e_{\alpha}(-\hat{\bm n}) e^{-2 \pi \nu i \hat{\bm n} \cdot {\bm r}/c} \mbox{,} \label{eq:incidentE}
\end{align}
where $\psi_{\alpha, \rm in}(\nu, \hat{\bm n})$ satisfies
\begin{align}
    \langle \psi^{\ast}_{\alpha, \rm in}(\nu_n, \hat{\bm n}) \psi_{\beta, \rm in}(\nu_m, \hat{\bm n}^\prime) \rangle & \approx k_{\rm B}T_{\rm s}(\hat{\bm n}) \delta_{\alpha \beta} \delta_{m n} \delta( \hat{\bm n} - \hat{\bm n}^\prime ) \mbox{.} \label{eq:rjcont}
\end{align}
We substitute Equations \eqref{eq:incidentE} and \eqref{eq:rjcont} for the sky-sourced electric fields into Equation \eqref{eq:skycontrib1} and obtain
\begin{align}
~~~ & \!\!\!\!
  \langle \psi^\ast_{{\rm c}_1}(\nu) \psi_{{\rm c}_2}(\nu) \rangle \vert_{\rm sky} \notag \\
  & = \frac{\xi^2}{Z_{\rm c}} \sin^2{(\phi_{\rm C})} \frac{4\pi \nu^2}{c^3} \sum_{\alpha} \int {\rm d}\hat{\bm n} \ [\hat{\bm x} \cdot {\bm e}^\ast_{\alpha}(-\hat{\bm n})] [\hat{\bm x} \cdot {\bm e}_{\alpha}(-\hat{\bm n})] \notag \\
  & \hspace{100pt} \times k_{\rm B}T_{\rm s}(\hat{\bm n}) e^{2 \pi \nu i \hat{\bm n} \cdot {\bm r}_{1 2}/c} \mbox{.}
\end{align}
We assume a monopole sky and define spherical angles with respect to the dipole separation, which we take to be along $\hat{\bm z}$. We define the azimuthal angle $\phi$ by $\hat{\bm n} \cdot \hat{\bm x} = \sin{\theta} \cos{\phi}$. We simplify as follows
\begin{align}
~~~ & \!\!\!\!
  \langle \psi^\ast_{{\rm c}_1}(\nu) \psi_{{\rm c}_2}(\nu) \rangle \vert_{\rm sky} \notag \\
  & = \sin^2{(\phi_{\rm C})} \frac{\xi^2 4\pi \nu^2 k_{\rm B}T_{\rm s}}{Z_{\rm c} c^3} \int {\rm d}\theta {\rm d}\phi \sin{\theta} (1 - \sin^2{\theta} \cos^2{\phi}) \notag \\
  & \hspace{100pt} \times e^{2 \pi i \nu r_{1 2} \vert \cos{\theta}/c} \notag \\
  & = \sin^2{(\phi_{\rm C})} \frac{(k \xi)^2 k_{\rm B}T_{\rm s}}{Z_{\rm c} c} \int {\rm d}\mu (1 + \mu^2) e^{i k r_{1 2} \mu} \mbox{.}
\end{align}

In going from the first line to the second, we substituted $k = 2\pi\nu/c$. We can evaluate the integral analytically. It is most instructive to express the result in terms of the sensitivity of the cross-correlation to the sky temperature as follows 
\begin{multline}
  \frac{1}{k_{\rm B}} \frac{\partial}{\partial T_{\rm s}} \langle \psi^\ast_{{\rm c}_1}(\nu) \psi_{{\rm c}_2}(\nu) \rangle = \frac{\sin^2{(\phi_{\rm C})}}{\pi} (k \xi)^2 \frac{Z_0}{Z_{\rm c}} \\
  \\[2pt]
  \times \frac{ k r_{12} \cos{(k r_{12})} + [(k r_{12})^2 - 1] \sin{(k r_{12})} }{ (k r_{12})^3 } \mbox{,} \label{eq:skysensitivity}
\end{multline}
where we used the fact that in c.g.s. units the impedance of free space is $Z_0 = 4\pi/c$. 

Using Equations (\ref{eq:selfrefl}) and (\ref{eq:crossrefl}), the right-hand side of Equation \eqref{eq:theoremstate} evaluates to
\begin{align}
  & -[U({\rm c}_1;{\rm c}_1)^\ast U({\rm c}_2;{\rm c}_1) + U({\rm c}_1;{\rm c}_2)^\ast U({\rm c}_2;{\rm c}_2)] \nonumber \\
  &~~~ = \frac{2\xi^2}{\pi\nu Z_{\rm c}} \sin^2(\phi_{\rm C}) \nonumber \\
  &~~~~~~ \times \frac{[(kr_{12})^2-1]\sin(kr_{12}) + kr_{12}\cos(kr_{12})}{r_{12}^3} \mbox{.}
\end{align}
This is identical to Equation \eqref{eq:skysensitivity}, as required by our theorem.

The total noise contribution to the cross-correlation in Equation \eqref{eq:netcorrelation} is
\begin{align}
  \langle \psi^\ast_{{\rm c}_1}(\nu) \psi_{{\rm c}_2}(\nu) \rangle \vert_{\rm noise} 
  & =  \langle \psi^\ast_{{\rm c}_1, {\rm in}}(\nu) \psi_{{\rm c}_2, {\rm out}}(\nu) \rangle \vert_{\rm noise} \nonumber \\
  &~~~ + \langle \psi^\ast_{{\rm c}_1, {\rm out}}(\nu) \psi_{{\rm c}_2, {\rm in}}(\nu) \rangle \vert_{\rm noise} \nonumber \\
  &~~~ + \langle \psi^\ast_{{\rm c}_1, {\rm out}}(\nu) \psi_{{\rm c}_2, {\rm out}}(\nu) \rangle \vert_{\rm noise} \notag \\
  & = \Bigl[ U({\rm c}_2; {\rm c}_1) + U({\rm c}_1; {\rm c}_2)^\ast \nonumber \\
  &~~~ + U({\rm c}_1; {\rm c}_1)^\ast U({\rm c}_2; {\rm c}_1) \nonumber \\
  &~~~ + U({\rm c}_1; {\rm c}_2)^\ast U({\rm c}_2; {\rm c}_2) \Bigr] k_{\rm B} T_{\rm n} \mbox{,}
\end{align}
where we used \eqref{eq:scatteringmat} for the relation between the output and input waveforms, and the relation in Equation \eqref{eq:noisetemp} for the input waveforms in the cables. 
We define the sensitivity to the noise temperature and readout the coefficients from Equations \eqref{eq:selfrefl} and \eqref{eq:crossrefl} to obtain
\begin{align}
  & \frac{1}{k_{\rm B}} \frac{\partial}{\partial T_{\rm n}} \langle \psi^\ast_{{\rm c}_1}(\nu) \psi_{{\rm c}_2}(\nu) \rangle \nonumber \\
  &~~~ = U({\rm c}_2; {\rm c}_1) + U({\rm c}_1; {\rm c}_2)^\ast + U({\rm c}_1; {\rm c}_1)^\ast U({\rm c}_2; {\rm c}_1) \nonumber \\
  &~~~~~~ + U({\rm c}_1; {\rm c}_2)^\ast U({\rm c}_2; {\rm c}_2) \nonumber \\
  &~~~ = - \frac{\sin^2{(\phi_{\rm C})}}{\pi} (k \xi)^2 \frac{Z_0}{Z_{\rm c}} \times \biggl[ {\rm Im} \left\{ e^{i(2 \vert \gamma \vert l + 2 \phi_{\rm C} + k r_{1 2})} \right. \nonumber \\
  &~~~~~~ \left. \times \frac{ (k r_{1 2})^2 + i k r_{1 2} - 1 }{(k r_{1 2})^3} \right\} \nonumber \\
  &~~~~~~ + \frac{ k r_{12} \cos{(k r_{12})} + [(k r_{12})^2 - 1] \sin{(k r_{12})} }{ (k r_{12})^3 } \biggr]
  \mbox{.} \label{eq:noisesens2}
\end{align}
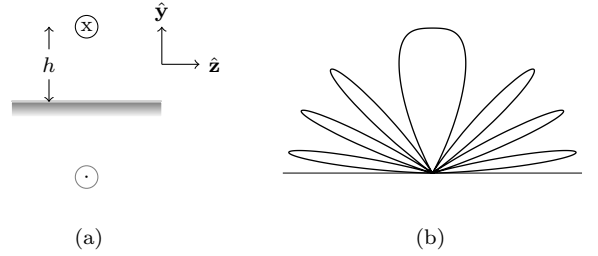
\begin{figure}[t]
  \centering
  \subfigure[]{
    \begin{tikzpicture}
      \draw[thick] (-1,0) -- (1,0);
      \shadedraw[white,shading=axis] (1,0) rectangle (-1,-0.2);
      \draw (0,1) circle (0.15);
      \node at (0,1) {x};
      \draw[<-] (-0.5,1) -- (-0.5,0.7);
      \node at (-0.5,0.5) {$h$};
      \draw[->] (-0.5,0.3) -- (-0.5,0);
      \draw[->] (1,0.5) -- (1.5,0.5);
      \node at (1.7,0.5) {$\hat{\bm z}$};
      \node at (-1.7,0.5) {};      
      \draw[->] (1,0.5) -- (1,1);      
      \node at (1,1.2) {$\hat{\bm y}$};      
      \draw[gray] (0,-1) circle (0.15);
      \node at (0,-1) {$\cdot$};
      \node at (0,-1.5) {};
    \end{tikzpicture}
    \label{fig:dipoleplane}
  }
  \hspace{60pt}
  \subfigure[]{
    \begin{tikzpicture}
      \node at (0,0) {};
      \put(-56,0){\includegraphics[width=4cm]{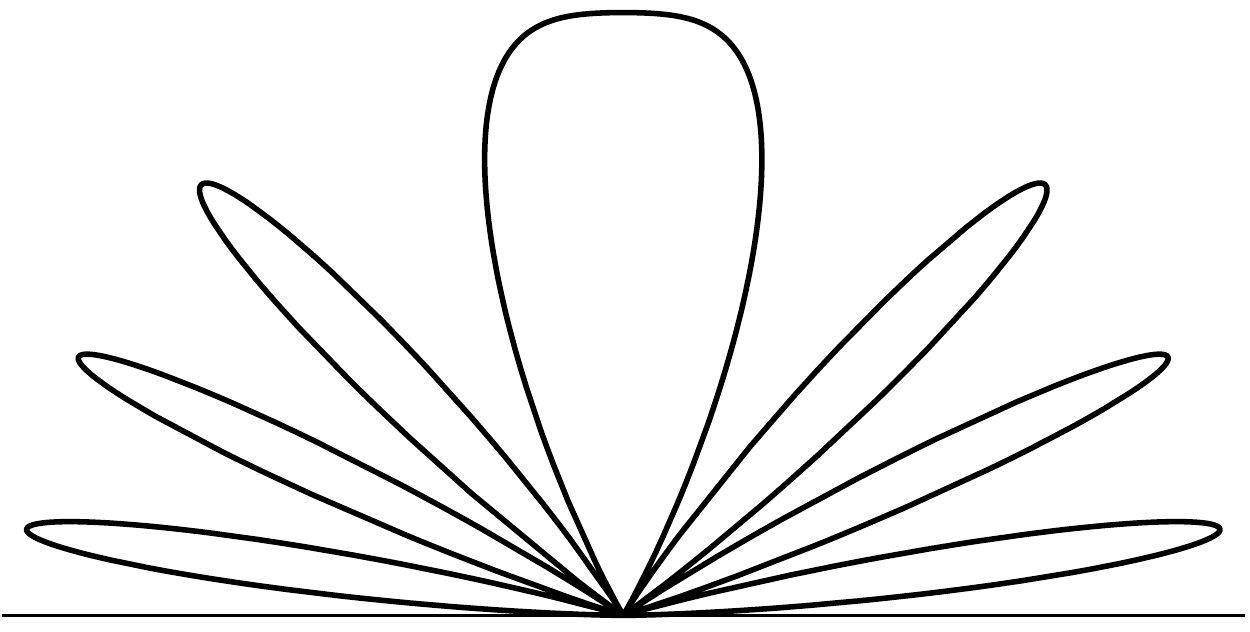}};
      \node at (0,-0.5) {};
    \end{tikzpicture}
    \label{fig:planebeampattern}
  }
  \hspace{50pt}
  \caption{\label{fig:reflectingdip} Panels (a) and (b) show a short dipole above a perfectly reflecting ground and the resulting far-field radiation pattern for a vertical displacement $h = 1.75 c/\nu$. The direction of the arrows in panel (a) shows the direction of the current densities in the dipole and its image, respectively.}
\end{figure}
In the last expression, the first (second) term is the sum of the first (last) two terms in the first line, and represents the correlation between the ingoing (reflected) Johnson noise at the first antenna's load with the received signal at the second antenna's load. We observe that the sensitivity in Equation \eqref{eq:skysensitivity} is fundamentally related to the size of the second term. This fact, which we found via calculation in this specific example, is a consequence of the general theorem of Section \ref{sec:proof}. 

\subsection{Dipoles above a reflecting ground}
\label{subsec:groundplane}

In a realistic setup the antennas are above a ground plane and additional beam-forming elements that  restrict the field of view, unlike the hypothetical scenario of two dipoles in free space. The theorem in Section \ref{sec:proof} does not rely on the whole sky being visible, so we expect the conclusions to hold regardless of the field of view. 

To demonstrate this, we consider a scenario with two short dipoles above a perfectly reflecting ground, with the dipole moments parallel to the ground. Figures \ref{fig:dipoleplane} and \ref{fig:planebeampattern} show the geometry of a single dipole and the resulting far-field radiation pattern, respectively.

We can check by inspection that Equations (\ref{eq:selfrefl})--(\ref{eq:dipoletransf}) and (\ref{eq:outunsimpl}) are unchanged. The only modifications are to the electric fields ${\bf E}_i(\nu)$: ground reflections modify both the sky and broadcast noise, and occult the lower hemisphere of the sky.

We write an expression for the sky-sourced electric field by considering the phase shift between the incident and reflected rays. We write this in a coordinate system oriented as in Figure \ref{fig:dipoleplane}, with the origin located on the ground plane.
\begin{align}
  {\bm E}_{\rm sky}({\bm r}, \nu) 
  & = i \sqrt{\frac{4\pi \nu^2}{c^3}} \sum_{\alpha} \int_{\hat{\bm n} \cdot \hat{\bm y}>0 } {\rm d}\hat{\bm n} \ \psi_{\alpha, \rm in}(\nu, \hat{\bm n}) 
  \notag \\ & 
  \!\!\!\!\!\!\!\!\times
  \left[ \bm e_{\alpha}(-\hat{\bm n}) e^{-2 \pi \nu i \hat{\bm n} \cdot {\bm r}/c} \right.  \left. - \bm e_{\alpha}^{(y)}(-\hat{\bm n}) e^{-2 \pi \nu i \hat{\bm n} \cdot {\bm r}^{(y)}/c} \right] \mbox{,} \label{eq:incidentErefl}
\end{align}
where we used the notation ${\bm a}^{(y)}$ to denote the vector ${\bm a}$ with the component normal to the plane (along $\hat{\bm y}$) reversed (i.e., ${\bm a}^{(y)} = {\bm a} - 2 ({\bm a} \cdot \hat{\bm y}) \hat{\bm y}$).
\begin{figure}[t]
  \begin{center}
    \includegraphics[width=\columnwidth]{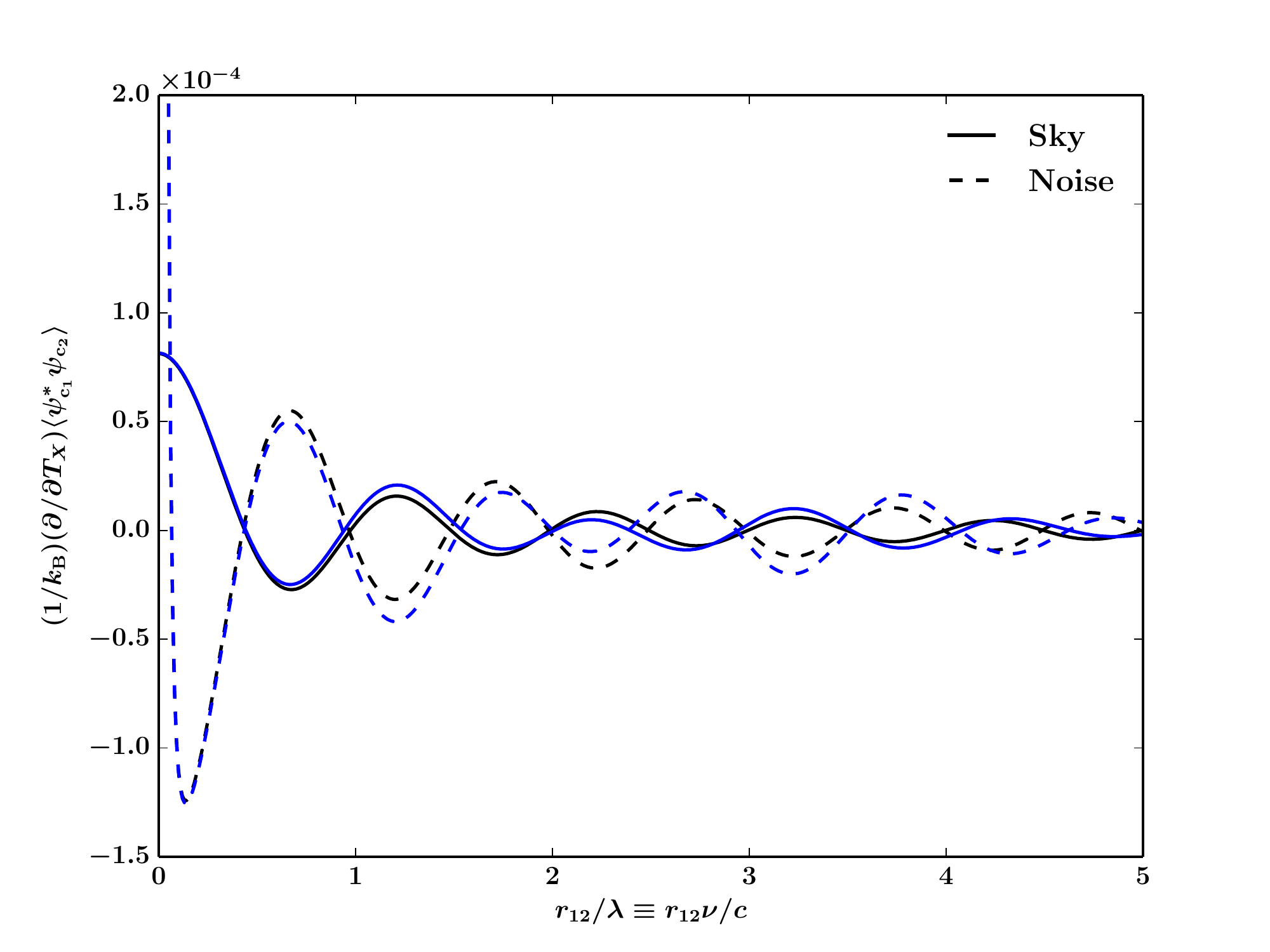}
  \caption{\label{fig:crosstalk} Solid and dashed lines show the sensitivity of the cross-correlation of the outputs of two short dipoles to the sky and noise temperatures, respectively, as a function of the separation in wavelength units. The black and blue lines differentiate results for the dipoles suspended in free-space and $h = 1.75 \lambda$ above a perfectly reflecting ground, respectively. The two dipoles have a capacitance of $1$ pF and conversion factor $\xi = 5$ cm, and are parallel and side by side. The figure is plotted for a frequency of $150$ MHz, with coaxial cables having an impedance of $50 \Omega$ and length $5$ m.}
\end{center}
\end{figure}

The driven dipole's electric field at the second one's location is modified from Equation \eqref{eq:dipoleEperp} to incorporate the reflected dipole, which has the opposite moment:
\begin{multline}
  {\bm E}({\bm r}_i, \nu; {\rm c}_j)
   = {\bm p}_i \Bigl[ \frac{(k r_{i j})^2 + i kr_{i j} - 1}{r_{i j}^3} e^{i k r_{i j}}
   \\ \hspace{30pt} 
  - \frac{(k \tilde{r}_{i j})^2 + i k \tilde{r}_{i j} - 1}{\tilde{r}_{i j}^3} e^{i k \tilde{r}_{i j}} \Bigr] \mbox{,} \label{eq:dipoleEperprefl} 
\end{multline}
where $\tilde{\bm r}_{i j} \equiv {\bm r}_i - {\bm r}_j^{(y)}$.
The lowest-order part of the coefficient $U({\rm c}_i; {\rm c}_i)$ is unchanged from Equation \eqref{eq:selfrefl}. We read off the cross-talk coefficient $U({\rm c}_i; {\rm c}_j)$ by substituting the electric field from Equation \eqref{eq:dipoleEperprefl} into the dipole response of Equation \eqref{eq:outunsimpl}. To lowest order in the conversion factor,
\begin{multline}
  U({\rm c}_i; {\rm c}_j) = i \frac{\xi^2}{\pi \nu Z_{\rm c}} \sin^2{(\phi_{\rm C}}) e^{2 i (\vert \gamma \vert l + \phi_{\rm C})} \times \\
  \left[ \frac{ (k r_{ij})^2 + i k r_{ij} - 1 }{r_{ij}^3} e^{i k r_{ij}} - \frac{ (k \tilde{r}_{ij})^2 + i k \tilde{r}_{ij} - 1 }{\tilde{r}_{ij}^3} e^{i k \tilde{r}_{ij}} \right] \mbox{.}
\end{multline}
As earlier, we obtain the sky contribution to the cross-correlation through the relation between the $\hat{\bm z}$ components of the electric field at the locations of the two antennas (via Equation \eqref{eq:skycontrib1}). By explicit calculation, we verify that the correlations between the first term in Equation \eqref{eq:incidentErefl} (the direct rays) at both locations, along with those between the last term (the reflected rays), add up to the result in Equation \eqref{eq:skysensitivity}, while the cross-correlations between the direct and reflected rays give the same term with the opposite sign and $r_{ij} \rightarrow \tilde{r}_{ij}$. Thus, we have the following sensitivity to the sky temperature:
\begin{multline}
  \frac{1}{k_{\rm B}} \frac{\partial}{\partial T_{\rm s}} \langle \psi^\ast_{{\rm c}_1}(\nu) \psi_{{\rm c}_2}(\nu) \rangle = \frac{\sin^2{(\phi_{\rm C})}}{\pi} (k \xi)^2 \frac{Z_0}{Z_{\rm c}} \\
  \times \Bigl[ \frac{ k r_{12} \cos{(k r_{12})} + [(k r_{12})^2 - 1] \sin{(k r_{12})} }{ (k r_{12})^3 } \\
  - \frac{ k \tilde{r}_{12} \cos{(k \tilde{r}_{12})} + [(k \tilde{r}_{12})^2 - 1] \sin{(k \tilde{r}_{12})} }{ (k \tilde{r}_{12})^3 } \Bigr] \mbox{.} \label{eq:skysensitivityrefl}
\end{multline}
The same additions and replacements to Equation \eqref{eq:noisesens2} also give the noise contribution to the cross-correlation. Our main theorem, Equation \eqref{eq:theoremstate}, is again satisfied, with the only difference being that a term with $r_{12}$ replaced by $\tilde r_{12}$ is subtracted from both sides of the equation.

Figure \ref{fig:crosstalk} shows the sensitivities of the two toy setups to the sky and noise temperatures for a fiducial set of system parameters. The large sensitivity to the noise temperature when the dipoles approach each other originates in the correlation between incoming noise waves at one cable and outgoing ones at the other, and are due to unphysically large fields at the location of the short dipole itself. In a real setup the dashed curves are cut off at small separations (around the conversion factor, $\xi$).

\section{Interferometer with well-separated antennas with no cross-talk or losses}
\label{sec:separated}

The setup in PLP15 consists of two antennas separated by some distance $R$. They show that because the interferometric fringe pattern $e^{-2\pi i \nu{\bm R}\cdot\hat{\bm n}/c}$ does not average to zero over the sky, the isotropic sky emission leads to a nonzero correlation of the electric field at the two antennas. At first sight, it would appear that if the two antennas can be arranged with no cross-talk $U({\rm c}_1;{\rm c}_2)=0$ and no losses, then the interferometer in PLP15 would be sensitive to the monopole of the sky, but both the ``cross-talk'' and ``dissipation'' terms in Equation (\ref{eq:theoremstate}) would be zero. This section examines a limiting case of the PLP15 setup (large $R$) in more detail. The resolution of the apparent disagreement between our intuition for the PLP15 setup and the theorem is that the usual formula for interferometric visibilities does not take into account sky radiation that scatters (or diffracts) off of one antenna and then goes into the other. We show here that the leading ($\sim 1/R$) contributions of the monopole to the visibility due to (i) the fringe pattern not integrating to zero and (ii) scattered radiation cancel.

For simplicity in what follows, we take the two antennas to be identical and separated by a large distance $R$ in their far field (i.e. if the antennas have diameter $D$, we take $R\gg D^2/\lambda$). We place antenna 1 at the origin and antenna 2 at position ${\bm R}=-{\bm r}_{12}$. We take the two antennas to be in free space (i.e.\ no ground plane).

\subsection{Sky contribution to the visibility}

First, the correlation between the output amplitudes $\psi_{{\rm c}_i,{\rm out}}$ seen at the two antennas at frequency $\nu$ from a sky at temperature $T_{\rm s}$ is
\begin{equation}
V_{\rm sky} 
= \sum_{\alpha,a} U^{\rm isol}({\rm c}_1;\alpha,\hat{\bm n}_a)^\ast U^{\rm isol}({\rm c}_2;\alpha,\hat{\bm n}_a) k_{\rm B}T_{\rm s}.
\end{equation}
In this equation, we take the coupling matrix ${\bm U}^{\rm isol}$ for the two isolated antennas (i.e., we compute the signal at c$_1$ neglecting the presence of antenna 2, and vice versa; the influence of the antennas on each other will be incorporated later).
We take the sky pixels to have size $\Omega_a$, and define the beam function
$\Upsilon_\alpha(\hat{\bm n}_a) = \Omega_a^{-1/2} U^{\rm isol}({\rm c}_1;\alpha,\hat{\bm n}_a)$,
which is independent of pixel size because, in accordance with Equation (\ref{eq:Ainc729}), the incident vector potential from sky port $\alpha a$ scales as $\propto \Omega_a^{1/2} \psi_{\rm in}$. Since the antennas are identical, the response functions differ by a factor corresponding to the propagation delay between the two antennas:
\begin{equation}
U^{\rm isol}({\rm c}_2;\alpha,\hat{\bm n}_a) = e^{-2\pi i {\bm R}\cdot\hat{\bm n}_a/\lambda} U^{\rm isol}({\rm c}_1;\alpha,\hat{\bm n}_a)
\end{equation}
and the visibility is
\begin{equation}
V_{\rm sky} 
= k_{\rm B}T_{\rm s} \int d^2\hat{\bm n} \,\sum_\alpha|\Upsilon_\alpha(\hat{\bm n})|^2 e^{-2\pi i {\bm R}\cdot\hat{\bm n}/\lambda}.
\label{eq:V12-usual}
\end{equation}
This is the usual formula, and in general the integral is not zero. In the limit where $R$ is large, we may find the leading contribution. We place the $z$-axis along $\hat{\bm R}$ without loss of generality, and use the standard spherical coordinates for $\hat{\bm n}$ (colatitude $\theta$ with $\mu=\cos\theta$ and longitude $\phi$). We further define $F(\hat{\bm n}) = \sum_\alpha|\Upsilon_\alpha(\hat{\bm n})|^2$.
Then we can rewrite Equation (\ref{eq:V12-usual}) as
\begin{equation}
V_{\rm sky} 
= k_{\rm B}T_{\rm s} \int_{S^2} d^2\hat{\bm n} \,F(\hat{\bm n})\, e^{-2\pi i R\mu/\lambda}.
\label{eq:A0}
\end{equation}
Now if $F$ is slowly varying---in particular, if it varies only on angular scales $\gg\lambda/R$---we see that $e^{-2\pi i R\mu/\lambda}$ is a rapidly varying function of position, and the integrand will average to zero. The exceptions occur when the phase is stationary, that is, at the North and South Poles, $\hat{\bm n} = \pm \hat{\bm e}_z$, where $2\pi R\mu/\lambda$ attains its extremal values $\pm 2\pi R/\lambda$. This suggests that we may apply the method of stationary phase. In the vicinity of the North Pole, the phase can be Taylor-expanded to second order as
\begin{equation}
e^{-2\pi i R\mu/\lambda} \approx 
 e^{-2\pi iR/\lambda} e^{\pi i(R/\lambda)(\hat n_x^2+\hat n_y^2)}.
\end{equation}
Integrating this over $d\hat n_x\,d\hat n_y$ using the Gaussian integral formula gives the replacement:
\begin{equation}
\int d^2\hat{\bm n}\, e^{-2\pi i R\mu/\lambda} \rightarrow e^{-2\pi iR/\lambda} \frac{ i\lambda}{R}.
\end{equation}
Combining this with the similar result at the South Pole gives the approximation to Equation (\ref{eq:A0}):
\begin{equation}
\!\!\! V_{\rm sky} 
\approx  \frac{i\lambda k_{\rm B}T_{\rm s}}{R} \left[
e^{-2\pi iR/\lambda}F(\hat{\bm e}_3) - e^{2\pi iR/\lambda}F(-\hat{\bm e}_3)
\right].
\label{eq:A1}
\end{equation}

Before proceeding, we note that Equation (\ref{eq:A1}) can be derived from integration by parts: we turn Equation (\ref{eq:A0}) into an integral $\int_{-1}^1 d\mu\int_0^{2\pi}d\phi$, and then apply integration by parts over $\mu$:
\begin{eqnarray}
V_{\rm sky} &=& \frac{i\lambda k_{\rm B}T_{\rm s}}{2\pi R} \Biggl[
\Bigl.\int_0^{2\pi} d\phi \,F(\mu,\phi) e^{-2\pi i R\mu/\lambda}\Bigr|_{\mu=-1}^1
\nonumber \\ &&
- \int_{-1}^1 d\mu\int_0^{2\pi} d\phi \, \frac{\partial F(\mu,\phi)}{\partial\mu} e^{-2\pi i R\mu/\lambda}
\Biggr].
\end{eqnarray}
Repeated integration by parts gives an asymptotic series with successively higher powers of $R^{-1}$. The leading term is Equation (\ref{eq:A1}).\footnote{One might object that $F(\mu,\phi)$ is not an analytic function at $\mu=\pm 1$, thereby rendering the sequence of derivatives of $F$ ill behaved. The integration over $\phi$ closes this loophole, because it guarantees that the longitude-averaged $F$ is an even function of $\theta$ and therefore can be expanded in even powers of $\theta$ (or $\pi-\theta$) and hence integer powers of $1-\mu$ (or $1+\mu$).}

Equation~(\ref{eq:A1}) demonstrates that for an interferometer with a long baseline (large $R$) and identical perfectly coupled antennas, the global sky contribution to the visibility is related to the antenna response in the directions $\pm\hat{\bm R}$ (i.e., along the antenna separation vector). In other words, for the global sky contribution to be nonzero (to leading order, $1/R$), either antenna 1 must ``see'' antenna 2, or 2 must see 1, or both.

\subsection{Scattering off of the antennas}

The fact that the sky contribution to the visibility is nonzero only when the antennas ``see'' each other suggests that we should consider scattered radiation from one antenna into the other (e.g. sky$\,\rightarrow 2\rightarrow 1$). One might think that this contribution declines as $R$ increases, but because in three dimensions the amplitude of a wave declines as the inverse of radius, the contribution of this pathway to the visibility is also $\propto 1/R$.

We begin by formulating the ``no cross-talk'' condition in terms of the beam function. The cross-talk between the antennas is proportional to the amplitude for antenna 1 to radiate in direction $\hat{\bm R}$ (toward antenna 2), there is a propagation amplitude, and then for antenna 2 to absorb radiation from direction $-\hat{\bm R}$ (from antenna 1). The overall amplitude for the cross-talk is then proportional to $\sum_\alpha \Upsilon_\alpha(\hat{\bm R})\Upsilon_\alpha(-\hat{\bm R})/R$, where we use the convention that the choice of polarization basis is the same in the $\hat{\bm R}$ and in the $-\hat{\bm R}$ directions;  we used reciprocity to relate the transmitting and receiving beam functions. The ``no cross-talk'' condition then states that
\begin{equation}
\sum_\alpha \Upsilon_\alpha(\hat{\bm R})\Upsilon_\alpha(-\hat{\bm R}) = 0.
\label{eq:no-cross-talk}
\end{equation}

In computing the scattering, we consider the sky$\,\rightarrow 1\rightarrow 2$ pathway first (and the sky$\,\rightarrow 2\rightarrow 1$ pathway is similar). This contributes to the visibility because the scattered radiation seen at 2 is correlated with the direct sky contribution to the signal at receiver 1. The contribution is
\begin{eqnarray}
V_{\rm sc1} &=& \langle \psi_{{\rm c}_1,{\rm out}}^\ast({\rm sky}\rightarrow 1) \psi_{{\rm c}_2,{\rm out}}({\rm sky}\rightarrow 1\rightarrow 2) \rangle
\nonumber \\
&\approx& k_{\rm B}T_{\rm s} \int d^2\hat{\bm n}\sum_{\alpha\beta}\Upsilon_\alpha^\ast(\hat{\bm n}) 
\frac{f_{\alpha\beta}(\hat{\bm n},\hat{\bm R})}R
\Upsilon_\beta(-\hat{\bm R}) e^{2\pi iR/\lambda},
\nonumber \\ &&
\label{eq:Vscat1}
\end{eqnarray}
where the far-field approximation has been made, and $f$ is the bidirectional scattering amplitude (with units of length). This is defined so that when an antenna is illuminated with a plane wave with electric field $E_{\rm in}(0)$ (measured at the origin) in polarization $\alpha$ from direction $\hat{\bm n}$, the scattered radiation in direction $\hat{\bm n}'$ at radius $r$ (in the far field) in polarization $\beta$ is
\begin{equation}
E_{\rm out} = f_{\alpha\beta}(\hat{\bm n},\hat{\bm n}')\frac{E_{\rm in}(0) e^{2\pi ir/\lambda}}r.
\label{eq:Eout-scat}
\end{equation}
We define the scattering amplitude $f$ with the boundary condition that radiation that couples into antenna 1 and travels down the coaxial cable sees an absorbing boundary condition.

Next, we note that the bidirectional scattering amplitude of a lossless antenna is not arbitrary, but is related to the antenna beam pattern and is constrained by the no cross-talk condition. This approach is similar in spirit to the derivation of the optical theorem \citep[e.g.][Equation (10.139), but with a cable present as well]{Jackson}.
Imagine a situation with an incident electromagnetic wave coming from direction $\hat{\bm R}$, with polarization vector $\hat{\boldsymbol\epsilon}$ and electric field amplitude $E_0$ (i.e., ${\bm E}_{\rm in} = E_0 \hat{\boldsymbol\epsilon} e^{-2\pi i\hat{\bm R}\cdot{\bm x}/\lambda}$). We further suppose that a signal $\psi_1$ is sent into the cable connected to antenna 1, and imagine for the present purposes that antenna 2 has been removed. At large distances $r$ from antenna 1 in direction $\hat{\bm n}$, the resulting electric field is the superposition of the incident, scattered, and transmitted waves:
\begin{eqnarray}
E_\alpha \,&=&\,
E_0\epsilon_\alpha e^{-2\pi i r\hat{\bm R}\cdot\hat{\bm n}/\lambda}
\nonumber \\ &&\,
+ \frac{e^{2\pi ir/\lambda}}{r} [ E_0 f_{\alpha\beta}(\hat{\bm n},\hat{\bm R})\epsilon_\beta + {\cal C}\psi_1\Upsilon_\alpha(\hat{\bm n})],
\end{eqnarray}
where ${\cal C}$ is a combination of constants, and we used reciprocity to write $f_{\alpha\beta}(\hat{\bm n},\hat{\bm R})$ where Equation (\ref{eq:Eout-scat}) would suggest $f_{\beta\alpha}(\hat{\bm R},\hat{\bm n})$. Now the net outgoing power is
\begin{equation}
P = \frac{c}{8\pi}\int_{S^2} |{\bm E}_{\rm out}|^2\,r^2\,d^2\hat{\bm n},
\end{equation}
again in the limit of large $r$, and where only the outgoing radiation (i.e.\ with $e^{2\pi ir/\lambda}$ instead of $e^{-2\pi ir/\lambda}$ dependence) is counted. This power has terms proportional to $|E_0|^2$, $|\psi_1|^2$, and a cross-term involving $E_0^\ast\psi_1$:
\begin{eqnarray}
P_{\rm cross} &=& \frac{c}{4\pi}\Re \Bigl\{ {\cal C}E_0^\ast\psi_1 \int_{S^2}
\Bigl[ \sum_\alpha
r e^{2\pi i r(1+\hat{\bm R}\cdot\hat{\bm n})/\lambda} \epsilon_\alpha^\ast \Upsilon_\alpha(\hat{\bm n})|_{\rm out}
\nonumber \\ &&~
+ \sum_{\alpha\beta} f^\ast_{\alpha\beta}(\hat{\bm n},\hat{\bm R})\epsilon^\ast_\beta \Upsilon_\alpha(\hat{\bm n})
\Bigr]\,d^2\hat{\bm n} \Bigr\},
\end{eqnarray}
where the ``out'' subscript indicates that only the portion of the integral where the radiation is outgoing is included. The first integral can be performed using the method of stationary phase to see that there is a contribution at $\hat{\bm R}\cdot\hat{\bm n}=\pm1$; only the $-1$ sign is outgoing. This leads to
\begin{eqnarray}
P_{\rm cross} &=& \frac{c}{4\pi}\Re \Bigl\{ {\cal C}E_0^\ast\psi_1 \Bigl[ \sum_\alpha
i\lambda \epsilon_\alpha^\ast \Upsilon_\alpha(-\hat{\bm R})
\nonumber \\ &&~
+ \sum_{\alpha\beta} \int_{S^2} f^\ast_{\alpha\beta}(\hat{\bm n},\hat{\bm R})\epsilon^\ast_\beta \Upsilon_\alpha(\hat{\bm n})
\,d^2\hat{\bm n} \Bigr] \Bigr\}.
\label{eq:temp-cross}
\end{eqnarray}
If the quantity in square brackets in Equation (\ref{eq:temp-cross}) is nonzero, the relative phase of $E_0$ and $\psi_1$ affects the amount of power radiated by the system, even though the amount of incident power and the amount of power sent into the cable depend only on $|E_0|$ and $|\psi_1|$. This is not necessarily a problem because the cable can carry away power (due to both the received signal $\propto E_0$ and the reflected signal $\propto \psi_1$). However, in the special case that the incident polarization is chosen to have $\sum_\alpha \epsilon_\alpha \Upsilon_\alpha(\hat{\bf R}) = 0$, the incident electromagnetic wave does not couple into the antenna (it has the ``wrong'' polarization), so the outgoing power in the cable is independent of $E_0$. In this case, the quantity in square brackets in Equation (\ref{eq:temp-cross}) must vanish. Because of the no cross-talk condition (Equation (\ref{eq:no-cross-talk})), this situation is realized for $\epsilon_\alpha = \Upsilon_\alpha(-\hat{\bf R})$. We therefore conclude that
\begin{eqnarray}
-i\lambda \sum_\alpha |\Upsilon_\alpha(-\hat{\bm R})|^2
&=& \sum_{\alpha\beta} \int_{S^2} f^\ast_{\alpha\beta}(\hat{\bm n},\hat{\bm R})\Upsilon_\beta^\ast(-\hat{\bm R})
\nonumber \\ && \times
 \Upsilon_\alpha(\hat{\bm n})
\,d^2\hat{\bm n}.
\label{eq:master-R}
\end{eqnarray}

The relation of Equation (\ref{eq:master-R}), substituted into Equation (\ref{eq:Vscat1}), gives
\begin{equation}
V_{\rm sc1} \approx \frac{i \lambda k_{\rm B}T_{\rm s}}{R} \sum_{\beta}
|\Upsilon_\beta(-\hat{\bm R})|^2 e^{2\pi i R/\lambda}.
\label{eq:Vscat1-sub}
\end{equation}
This cancels one of the terms in Equation (\ref{eq:A1}). The sky$\,\rightarrow 2\rightarrow 1$ pathway can be calculated similarly and cancels the other term. Thus we see that, to order $1/R$, the combined visibility is
\begin{equation}
V_{\rm sky} + V_{\rm sc1} + V_{\rm sc2} = 0.
\end{equation}
Thus, when we consider two identical well-separated antennas with zero cross-talk and zero loss, the visibility obtained from the traditional interferometer fringe pattern integrated against a monopole from the sky is nonzero. However, the nonzero contribution is dominated by the regions of the sky where the interferometer phase is stationary, $\hat{\bm n} = \pm \hat{\bm R}$ -- exactly the regions of the sky where antenna 1 obstructs the view from antenna 2, or vice versa. When the radiation scattered from one antenna into the other is taken into account, the total visibility (which is what would be observed in a real interferometer) cancels out and the setup is not sensitive to the global sky signal.

As a final comment, one might think that this scattering problem can be circumvented by making the antennas very small (i.e., of size $\ll \lambda$) and not resonant at the frequencies of interest, so that their scattering cross sections are negligible. This does work: this is the short-dipole problem outlined in Section \ref{sec:shortdipole}, where the sensitivity of the interferometer to an isotropic sky temperature is nonzero and declines as $\propto 1/R$ ($r_{12}=R$) at large separation. Of course, the cross-talk would then be nonzero because the short dipole lacks the directionality to make the antenna sensitive in the $\hat{\bm R}$ direction (to make the stationary-phase approximation to $V_{\rm sky}$ nonzero) while simultaneously avoiding sensitivity in the $-\hat{\bm R}$ direction (required to prevent cross-talk).

\section{Conclusions and Discussion}
\label{sec:discussion}

The redshifted $21$ cm radiation background is an important probe of cosmological recombination and the preceding cosmic dark ages. In particular, the global or sky-averaged signal contains information about the thermal history of neutral hydrogen in the early universe. However, this is a challenging measurement to make because of large local foregrounds, as well as the difficulty of calibrating the receivers' noise properties. In particular, the latter difficulty arises in any setup that uses the autocorrelation of the waveform measured by a receiver attached to a single antenna. Motivated by these challenges, several groups have recently proposed innovative methods to measure the global signal using interferometric setups. In these methods, the measured quantity is not the autocorrelation, but the cross-correlation of the waveforms from different antennas. 

In this paper we study the physical principles underpinning the response of an arbitrary multiple-readout channel setup to uniform radiation in its field of view. We visualize a readout channel as the terminus of a coaxial cable connecting to an idealized amplifier. The cross-correlation of the signals in two readout channels gives the usual visibilities of radio astronomy. We argue that such cross-correlations are sensitive to a global signal only if at least one of the following three conditions is satisfied: (1) There is a nonzero amount of cross-talk between the two readout channels, and when these channels are locally driven at least one of them sees a nonzero reflected power. (2) There are other channels that exhibit cross-talk with both the channels. (3) There are dissipative elements within the setup that can send noise into both the channels. Moreover, in the first two cases the sensitivity to a global signal is directly related to the cross-talk and in the latter it is related to the correlated input noise, both of which introduce a systematic noise bias. We illustrated these results in two setups involving short dipole antennas, as well as one with an interferometer with no cross-talk. Hence the system has a similar response (in terms of magnitude) to a global sky temperature and a local noise temperature. 

In conventional interferometric setups the local noise contribution to the visibility is reduced by minimizing the cross-talk between the elements. The results in this paper imply that any such reduction (in our examples, this is accomplished by a physical separation) is accompanied by a similar one in the sensitivity to the global signal. Hence, for any setup that aims to perform such a measurement, the same considerations govern both the systematic noise bias and the sensitivity to the signal. Any interferometric setup aiming to measure the global signal must carefully study its noise properties to understand its sensitivity.

While these results are sobering, we make no attempt to judge the relative value of having interferometric setups vis-a-vis conventional single element ones (indeed, our results blur the line between them). In particular, the issue of noise bias does not preclude the former, for the same reason as for the latter. We also observe that, at least in principle, the thermal noise bias can be reduced by cooling the setup, or characterized by varying the temperatures of the relevant elements. We leave any detailed considerations (such as those of strong frequency dependence) of realistic designs for future work. 

\begin{acknowledgments}
We would like to thank Michael Eastwood, Shriharsh Tendulkar, and Kris Sigurdson for several helpful discussions, and Matias Zaldarriaga for comments on the manuscript. We also thank Aaron Parsons, Adrian Liu, and Morgan Presley for useful feedback at an early stage of this work.

We thank the anonymous referee for useful comments, and in particular for asking the question that inspired Section \ref{sec:separated}.

TV gratefully acknowledges support from the Schmidt Fellowship and the Fund for Memberships in Natural Sciences at the Institute for Advanced Study. CH is supported by the US Department of Energy, the David \& Lucile Packard Foundation, and the Simons Foundation. T.-C. C. acknowledges support from MoST grant 103-2112-M-001-002-MY3. Part of the research described in this paper was carried out at the Jet Propulsion Laboratory, California Institute of Technology, under a contract with the National Aeronautics and Space Administration.
\end{acknowledgments}

\appendix
 
\section{Discretizing the sky}
\label{sec:discrete}
\begin{figure}[t]
\center{\includegraphics[width=\columnwidth]{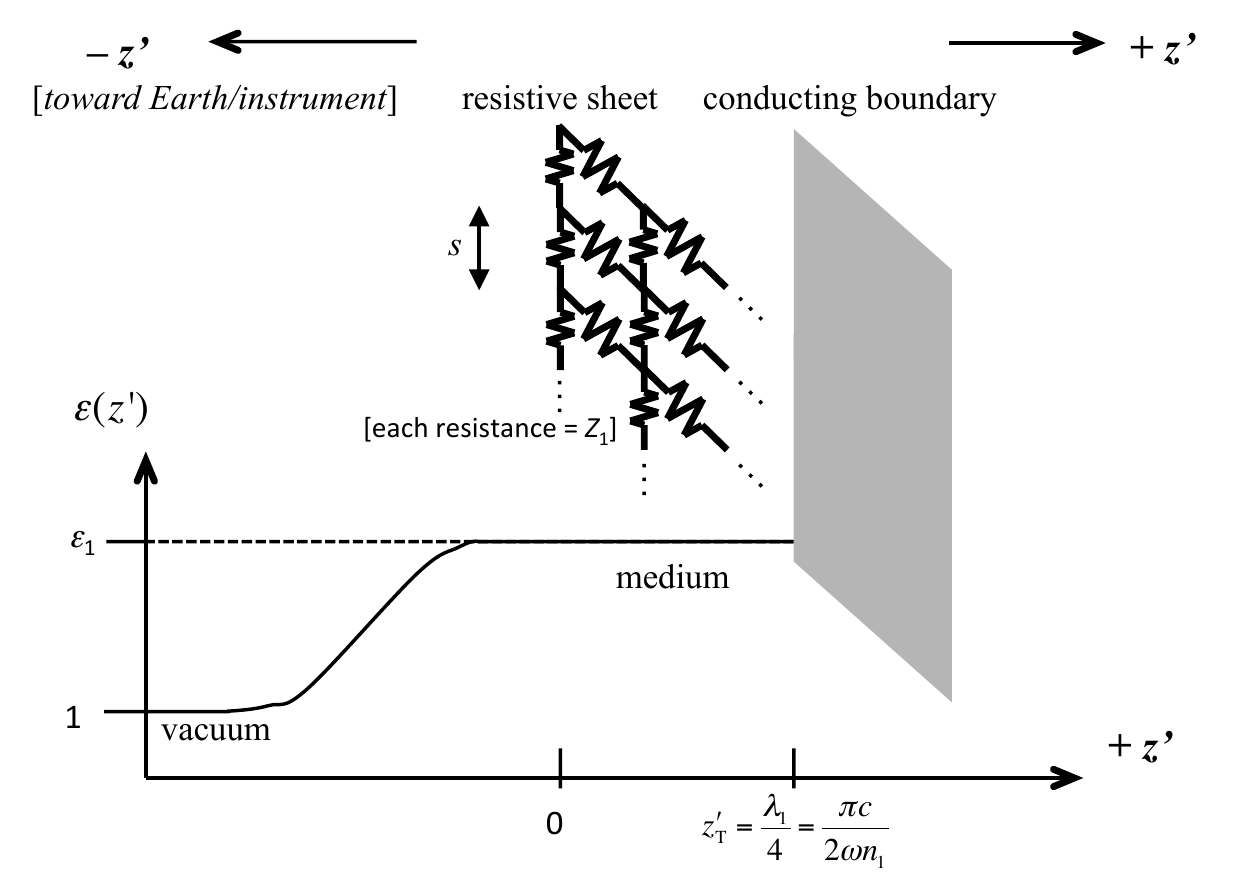}}
\caption{\label{fig:discrete}Diagram of the absorbing sky-layer described in the main text. The resistive sheet is located a distance ${\cal R}$ from the observer and is the origin of a local (primed) coordinate system (i.e., it defines $z'=0$). It is embedded in a high-permittivity dielectric, $\epsilon_1\gg 1$, that adiabatically tapers to vacuum on the left (inward-facing) side and is terminated by a conducting boundary at the right (outward-facing) side. This diagram is not to scale.}
\end{figure}

In this section we describe a scheme for defining discrete modes of the EM field coming in from the sky, in such a way as to make contact with the network picture described in the main text. The key is to replace the infinitely distant sky (where radiation can travel in and out) with a set of network ports. We describe the geometry of our scheme for a single frequency, $\nu$ (or equivalently, angular frequency $\omega = 2\pi \nu$).

Imagine replacing the sky with an absorbing spherical shell of very large radius ${\cal R}$ far from the experiment. (We imagine taking the limit as ${\cal R}\rightarrow\infty$, so in the case of a ground-based experiment this shell surrounds the whole Earth.) A concrete realization of this shell is the geometry shown in Figure~\ref{fig:discrete}. This geometry has a lossless material of graded real dielectric constant $\epsilon$ that rises adiabatically from 1 (vacuum) to $\epsilon_1$. In this medium, the index of refraction is $n_1=\epsilon_1^{1/2}$; we take the limit as $\epsilon_1$ becomes large (to refract all incident radiation so that its direction of propagation is near normal when it hits the termination described below). We then embed a square-mesh grid of resistors, each of resistance $Z_1 = Z_0/n_1 = 4\pi /n_1 c$ ($Z_0$ is the impedance of free space), and a quarter-wavelength beyond that we place a perfectly conducting sheet. In the local coordinate system in the vicinity of these sheets where $z'$ points away from the experiment, we thus have the resistor grid at $z'=0$ and the conducting sheet at $z'=z'_{\rm T} = c \pi/ (2 \omega n_1)$. To appear continuous, the resistor grid must have unit cell size $s\ll \lambdabar = c/(kn_1)$, where $k$ is the radiation's wavenumber in free space. The conducting sheet -- which is globally a spherical conducting shell enclosing the whole system -- ensures that no energy leaks out: all electromagnetic wave energy is ultimately either absorbed in the resisting sheet or enters the coaxial cables in the experiment.

The above spherical shell can be shown by standard techniques to be perfectly absorbing in the limit that $\epsilon_1$ is large; we sketch the proof here. The graded dielectric constant refracts radiation at angle of incidence $\theta_{\rm i}$ to an angle $\theta_{\rm f}$ given by Snell's law, $\sin\theta_{\rm f} = (\sin\theta_{\rm i})/n_1<1/n_1$. For a wave propagating in the $z'$-direction (i.e. fields independent of $x'$ and $y'$) and polarized in the $x'$-direction, Maxwell's equations give
\begin{equation}
\frac{\partial E'_x}{\partial z'} = \frac{i\omega}cB'_y ~~~{\rm and}~~~
\frac{\partial B'_y}{\partial z'} = \frac{i\omega\epsilon_1}cE'_x,
\label{eq:wave}
\end{equation}
with the boundary condition that $E'_x=0$ at the conducting termination $z'=z'_{\rm T}$, and the jump conditions
\begin{align}
\lim_{z'\rightarrow 0^+} E'_x = \lim_{z'\rightarrow 0^-} E'_x
 ~~ &~ {\rm and}~~~ \lim_{z'\rightarrow 0^+} B'_y - \lim_{z'\rightarrow 0^-} B'_y \nonumber \\
= -\frac{4\pi}cK'_x & = -\frac{4\pi}{cZ_1} E'_x(z'=0),
\end{align}
where $K'_x$ is the linear current density (units: statamp cm$^{-1}$) in the resistive mesh. The wave equation gives the solutions $E'_x\propto e^{\pm ik_1z'}$, where $k_1=\omega n_1/c$ is the local wavenumber; we may thus write
\begin{equation}
E'_x = \left\{\begin{array}{lll} A_1 e^{ik_1z'} + A_2 e^{-ik_1z'} & & z'<0 \\ (A_1+A_2)\cos (k_1z') & & z'>0 \end{array}\right.,
\end{equation}
where at $z'>0$ the terminating boundary condition eliminates the $\sin (k_1z')$ solution, and the continuity of $E'_x$ implies that the coefficient is $A_1+A_2$. The jump condition on $B'_y$ and hence on the derivative of $E'_x$ then forces $A_2=0$.

As in the main text, we then replace each resistor with an infinitely long coaxial cable of impedance $Z_1$. The ``distant sky'' has now been replaced by a suite of cables, and thermal radiation from the sky is swapped out for resistor Johnson noise and ultimately thermal input noise coming down each cable.

We now have a setup with a suite of $N$ cables (which may be ``sky'' cables or physical cables attached to the antenna), all connected to a gigantic network. The voltage in the $I$th cable may be broken down into ``ingoing'' and ``outgoing'' modes,
\begin{equation}
V_I(x,t) = V_{I,{\rm in}}(t-x/v_I) + V_{I,{\rm out}}(t+x/v_I),
\label{eq:in.plus.out}
\end{equation}
where $v_I$ is the propagation velocity in cable $I$. Note that the sense of $x$ is that positive $x$ is toward the network and negative $x$ is away. We consider the behavior over a finite time ${\cal T}$ (which will be taken to $\infty$), so that all fields may be Fourier transformed as sums of modes with frequency $\nu_m = m/\cal T$ and spacing $\Delta\nu=1/\cal T$. Then we may write these modes as
\begin{align}
  & V_{I, {\rm in}/{\rm out}}^{(\cal T)}(t\mp x/v_I) = \sqrt{\frac{Z_I}{2\mathcal{T}}} \sum_{m>0} \nonumber \\
  & ~~~ \left[ \psi_{I, {\rm in}/{\rm out}}^{(\cal T)}(\nu_m) e^{ \pm i\vert \gamma \vert x - 2\pi i \nu_m t} + {\rm c.c.}\right] \mbox{,} \label{eq:cablemodes}
\end{align}
where $Z_{I}$ is the cable's characteristic impedance, $\gamma = i \vert \gamma \vert$ is the (lossless) propagation constant with $\vert\gamma\vert=2\pi\nu_m/v_I$, and the factor of $1/\sqrt{2}$ is chosen so that the modulus squared of the positive frequency component, $\psi_{I, \rm out}^{(\cal T)}$, averages to the one-sided power spectral density. That is, the time-averaged outgoing power is
\begin{equation}
P_{I,\rm out} = \sum_{m>0}  \left| \psi_{I, {\rm in}/{\rm out}}^{(\cal T)}(\nu_m) \right|^2 \Delta\nu.
\end{equation}

Now in the square-mesh grid there are resistors in two orthogonal directions in the plane of the sphere, which we denote horizontal (H) and vertical (V), and we can define direction indices $\alpha\in\{{\rm H},{\rm V}\}$. The grid has a unit cell solid angle $\Delta\Omega_a = s^2/{\cal R}^2$; we will use the index $a$ to denote the cells. There is a total of $N_{\rm sky} = 4\pi{\cal R}^2/s^2$ such cells, and hence $2N_{\rm sky}$ sky ports in the network.

Our next objective is to determine the field created in the experiment's vicinity by the signal entering the network at a sky port. If we consider that each sky port is a resistor in one of the sheets that we have placed on the sky (replaced by a cable), then the input signal in each such resistor given by
\begin{equation}
V_{\rm in}[{\rm s}(\alpha,\hat{\bf n}_a])
= \sqrt{\frac{Z_1}{2{\cal T}}} \sum_{m>0} \psi_{\rm in}[{\rm s}(\alpha,\hat{\bf n}_a,\nu_m)] e^{-2\pi i\nu_mt} + {\rm c.c.},
\label{eq:sky-in}
\end{equation}
where we placed the (arbitrary) $x=0$ point at the resistor itself. Recall that if a cable of impedance $Z_1$ is attached to a circuit, then an input voltage $V_{\rm in}$ has the same effect as a source of e.m.f. $2V_{\rm in}$ connected in series with a resistor of resistance $Z_1$. If we consider the input to a {\em single} resistor pointed in the $x'$-direction, this is equivalent to an externally applied electric field in the sheet of magnitude $2V_{\rm in}/s$ applied over a unit cell of area $s^2$, that is, with
\begin{equation}
\int E'_{x,\rm eff} \,dx'\,dy' = 2sV_{\rm in}.
\end{equation}
If we integrate Maxwell's equations over $x'$ and $y'$, then we once again find the wave equation
\begin{align}
\frac{\partial}{\partial z'}\int E'_x \,dx'\,dy' = & \frac{i\omega}c\int B'_y \,dx'\,dy' ~~~{\rm and} \nonumber \\ 
\frac{\partial}{\partial z'}\int B'_y \,dx'\,dy' & = \frac{i\omega\epsilon_1}c\int E'_x \,dx'\,dy',
\end{align}
but this time the jump condition at the mesh is given by
\begin{align}
& \lim_{z'\rightarrow 0^+} \int B'_y \,dx'\,dy' - \lim_{z'\rightarrow 0^-} \int B'_y \,dx'\,dy' \nonumber \\
&~~~ = -\frac{4\pi}c\int K'_x \,dx'\,dy' \nonumber \\
&~~~ = -\frac{4\pi}{cZ_1}\int E'_x(z'=0)\,dx'\,dy' - \frac{8\pi s}{cZ_1}V_{\rm in}.
\label{eq:der-int}
\end{align}
The solution to this is given by
\begin{equation}
\int E'_x\,dx'\,dy' = \left\{\begin{array}{lll} A_3 e^{-ik_1z'} & & z'<0 \\ A_3\cos (k_1z') & & z'>0 \end{array}\right.,
\end{equation}
where in the $z'<0$ regime we only have waves emitted toward the experiment (we are considering the propagation of radiation from only a single input port, so an outgoing-wave-only boundary condition is appropriate here), and the $A_3$ appears in both cases by continuity of $E'_x$. The derivative jump condition (Equation (\ref{eq:der-int})) requires
\begin{equation}
A_3 = -\frac{8\pi s V_{\rm in}}{c^2k_1\omega^{-1}Z_1 + 4\pi} = -sV_{\rm in},
\end{equation}
where we used the explicit formulae for $Z_1$ and $k_1$ to achieve simplification in the last step. Following the adiabatic (Wentzel-Kramers-Brillouin) solution to the wave equation at negative $z'$ gives in the vacuum region
\begin{equation}
\int E'_x\,dx'\,dy' = -\epsilon_1^{1/4} e^{i\Phi} sV_{\rm in} e^{-i\omega z'/c},
\label{eq:e-vac}
\end{equation}
where $\Phi$ is a phase shift that depends on the details of the transition from high to vacuum dielectric constant, and whose value we will not need. The Poynting flux of energy in the vacuum and dielectric regions is matched owing to the factor of $\epsilon_1^{1/4}$. 

From this property of the electric field in the vacuum region near the spherical shell's surface, we may obtain the electric field incident from this system at the position of the experimental setup, which is given by the diffraction integral \citep[e.g.][Equation (10.85)]{Jackson} in the far-field limit $kR\gg 1$ with $k=\omega/c$:
\begin{equation}
{\bf E}({\bf x}) = \frac{k}{2\pi i} \int_{\cal S} \frac{e^{ikR}}{R} {\bf E}(\tilde{\bf x})\,\hat{\bf n}'\cdot\hat{\bf R}\,d^2\tilde{\bf x},
\label{eq:temp1}
\end{equation}
where the integral is over the surface ${\cal S}$ at $z'=z'_{\cal S}$, $R$ is the distance from the surface to the point ${\bf x}$, and $\hat{\bf n}'$ is the normal to surface ${\cal S}$ (pointed toward the experiment). If ${\bf x}$ is near the origin $O$ of the experiment's coordinate system and satisfies $|{\bf x}|\ll\sqrt{\lambda{\cal R}}$ (so that the external sphere is large enough to be in the far field as seen from all points of interest) then we write $R = {\cal R} + z'_{\cal S} - {\bf x}\cdot\hat{\bf n}_a$, where $\hat{\bf n}_a$ is the direction from the experiment to cell $a$ on the resistive mesh. In this case, Equation (\ref{eq:temp1}) simplifies to
\begin{equation}
{\bf E}({\bf x}) = \frac{k}{2\pi i\cal R} e^{ik({\cal R} + z'_{\cal S})} e^{-ik{\bf x}\cdot\hat{\bf n}_a}
\int_{\cal S} {\bf E}(\tilde{\bf x})\,d^2\tilde{\bf x}.
\end{equation}
Then we use Equation (\ref{eq:e-vac}) and let $\hat{\bf e}_\alpha$ be the unit vector in the direction of the resistor being considered ($x'$ or $y'$), and obtain
\begin{align}
  {\bf E}({\bf x}) &= -\frac{k}{2\pi i\cal R} e^{-ik{\bf x}\cdot\hat{\bf n}_a}
  \epsilon_1^{1/4} e^{i(\Phi + k \cal R)} sV_{\rm in} \hat{\bf e}_\alpha
  \nonumber \\
  & =
  - \sqrt{\frac{Z_1}{2{\cal T}}} \sum_{m>0} 
  \frac{ks}{2\pi i\cal R} e^{-ik{\bf x}\cdot\hat{\bf n}_a}
  \epsilon_1^{1/4} e^{i(\Phi + k \cal R)}
  \psi_{\rm in} \nonumber \\
  &~~~ \times [{\rm s}(\alpha,\hat{\bf n}_a,\nu_m)] e^{-2\pi i\nu_mt} \hat{\bf e}_\alpha + {\rm c.c.} \nonumber \\
  & =
  \sum_{m>0} 
  \frac{ik\Omega_a^{1/2}}{\sqrt{2\pi c\cal T}} e^{-ik{\bf x}\cdot\hat{\bf n}_a}
  e^{i(\Phi + k \cal R)} \psi_{\rm in} \nonumber \\
  &~~~ \times [{\rm s}(\alpha,\hat{\bf n}_a,\nu_m)] e^{-2\pi i\nu_mt} \hat{\bf e}_\alpha + {\rm c.c.}.
  \label{eq:temp2}
\end{align}
If the electric field incident on the detector is described in the Coulomb gauge, where electromagnetic waves are described entirely with the vector potential, then we have ${\bf E} = -(1/c)\partial{\bf A}/\partial t = i(\omega/c){\bf A} = ik{\bf A}$ and hence:
\begin{align}
  {\bf A}_{\rm incident}({\bf x}) & = 
  \sum_{m>0} 
  \frac{\Omega_a^{1/2}}{\sqrt{2\pi c\cal T}}
  e^{i(\Phi + k \cal R)}
  \psi_{\rm in} \nonumber \\
  &~~~ \times [{\rm s}(\alpha,\hat{\bf n}_a,\nu_m)] e^{-2\pi i\nu_m(t+{\bf x}\cdot\hat{\bf n}_a/c)} \hat{\bf e}_\alpha + {\rm c.c.}.
  \label{eq:Ainc729}
\end{align}
This corresponds to Equation (\ref{eq:decomposition}) if we (i) recognize that the choice of polarization basis $\{\hat{\bf e}_\alpha\}$ depends on direction on the sky, (ii) absorb the phase factor of $\exp{(i k \cal R)}$ into the incoming mode amplitudes (i.e., use the short hand $\psi_{\alpha,\rm in}(\hat{\bf n}_a,\nu_m)]$ for $\exp{(i k \cal R)} \psi_{\rm in}[{\rm s}(\alpha,\hat{\bf n}_a,\nu_m)]$;), and (iii) choose $\Phi+k{\cal R}=0$---recall that $\Phi$ was the phase shift accumulated in the graded dielectric, which may be set to any value by choosing the function $\epsilon(z')$. Note that by absorbing the phase factor into the incoming mode's amplitude and setting the combination $\Phi+k{\cal R}$ to zero, we are effectively measuring its phase at the experiment's location, rather than the point of generation. To be consistent we have to measure the outgoing mode at the same location, which leads to its phase being redefined by a factor of $\exp{(-i k \cal R)}$. This opposite change in the incoming and outgoing modes' phases is characteristic of a transformation that preserves the reciprocity of the scattering matrix.

\section{Effects of cable termination}
\label{sec:termination}

The derivation presented in the body of the paper assumed that the coaxial cables at the readouts are connected to idealized amplifiers (with infinite input impedance), and impedance matched by appropriate resistive loads in parallel (see Figure \ref{fig:drivecircuit}). This is not a crucial requirement for the theorem presented in Section \ref{sec:proof}. In this section, we demonstrate this in a scenario with open termination at the readout channels. We treat the case of an isotropic sky to reduce the number of terms, but the inclusion of an anisotropy map $\Delta T_{\rm s}(\hat{\bm n}_a)$ would proceed in a manner similar to Section \ref{sec:proof}.

In this case, the incoming modes in the readout channels are not noise voltages satisfying Equation \ref{eq:noisetemp}, but rather are given by $\psi_{{\rm c}_i, {\rm in}} = \psi_{{\rm c}_i, {\rm out}}$ (the voltage reflection coefficient at an open termination is unity). The next step is to substitute this relation into Equation \eqref{eq:beamu} for the outgoing signals. It is convenient to subdivide the scattering matrix of Equation \eqref{eq:scatteringmat} into the readout channel to readout channel, other to other, and readout channel to other subblocks: $U_{\rm cc}$, $U_{\rm oo}$ and $U_{\rm co}$, respectively. The other channels run over the sky cables of Appendix \ref{sec:discrete} and the dissipative cables of Section \ref{sec:formalism} (the latter run over all the dissipative elements in the setup).

Then the relation between incoming and outgoing signals at the open termination(s) yields
\begin{align}
  \psi_{{\rm c}, {\rm in}} & = U_{\rm cc} \psi_{{\rm c}, {\rm in}} + U_{\rm co} \psi_{{\rm o}, {\rm in}} \mbox{,} \ {\rm i.e.,} \\ 
  \psi_{{\rm c}, {\rm in}} & = (\mathds{1} - U_{\rm cc} )^{-1} U_{\rm co} \psi_{{\rm o}, {\rm in}} \mbox{,}
\end{align}
where $\psi_{\rm c}$ and $\psi_{\rm o}$ are column vectors of the waveforms in the readout channels and the other channels, respectively.

The measured cross-correlations (visibilities) are the off-diagonal parts of the general covariance matrix, which is the following expectation value:
\begin{align}
  \langle \psi_{{\rm c}} \psi_{{\rm c}}^\dagger \rangle & = 4 \langle \psi_{{\rm c}, {\rm in}} \psi_{{\rm c}, {\rm in}}^\dagger \rangle \nonumber \\
  & = 4 \left\langle (\mathds{1} - U_{\rm cc} )^{-1} U_{\rm co} \psi_{{\rm o}, {\rm in}} \left[ (\mathds{1} - U_{\rm cc} )^{-1} U_{\rm co} \psi_{{\rm o}, {\rm in}} \right]^\dagger \right\rangle \nonumber \\
  & = 4 (\mathds{1} - U_{\rm cc} )^{-1} U_{\rm co} \left\langle \psi_{{\rm o}, {\rm in}} \psi_{{\rm o}, {\rm in}}^\dagger \right\rangle U_{\rm co}^\dagger (\mathds{1} - U_{\rm cc}^\dagger )^{-1} \mbox{,} \label{eq:matrixcovar}
\end{align}
where in the first line we have used the relation between incoming and outgoing signals at the open termination. We separate the expectation value on the RHS of Equation \eqref{eq:matrixcovar} (and the other channels) into the sky and dissipative cables, and assume a uniform sky temperature. 
\begin{align}
  \langle \psi_{{\rm c}} \psi_{{\rm c}}^\dagger \rangle & = 4 k_{\rm B} T_{\rm s} (\mathds{1} - U_{\rm cc} )^{-1} U_{\rm cs} U_{\rm cs}^\dagger (\mathds{1} - U_{\rm cc}^\dagger )^{-1} \nonumber \\
  &~~~+ 4 k_{\rm B} (\mathds{1} - U_{\rm cc} )^{-1} U_{\rm cd} T_{\rm d} U_{\rm cd}^\dagger (\mathds{1} - U_{\rm cc}^\dagger )^{-1} \mbox{.} \label{eq:aribtraryout}
\end{align}
Here, $T_{\rm d}$ is a diagonal matrix whose entries are the noise temperatures of all the dissipative elements in the setup. In the manner of the derivation in Section \ref{sec:proof}, we define the sensitivity to the uniform sky temperature as follows
\begin{align}
  & \frac{1}{k_{\rm B}} \frac{\partial}{\partial T_{\rm s}} \langle \psi_{{\rm c}} \psi_{{\rm c}}^\dagger \rangle 
  = 4 (\mathds{1} - U_{\rm cc} )^{-1} U_{\rm cs} U_{\rm cs}^\dagger (\mathds{1} - U_{\rm cc}^\dagger )^{-1} \nonumber \\
  &~~~ = 4 (\mathds{1} - U_{\rm cc} )^{-1} (\mathds{1} - U_{\rm cc} U_{\rm cc}^\dagger - U_{\rm cd} U_{\rm cd}^\dagger) (\mathds{1} - U_{\rm cc}^\dagger )^{-1} \nonumber \\
  &~~~ = 4 (\mathds{1} - U_{\rm cc} )^{-1} (\mathds{1} - U_{\rm cc} U_{\rm cc}^\dagger) (\mathds{1} - U_{\rm cc}^\dagger )^{-1} \nonumber \\
  &~~~~~~ - \sum_{{\rm d}_i} \frac{1}{k_{\rm B}} \frac{\partial}{\partial T_{{\rm d}_i}} \langle \psi_{{\rm c}} \psi_{{\rm c}}^\dagger \rangle \nonumber \\
  &~~~ = 4 (\mathds{1} - U_{\rm cc} )^{-1} \left[ \mathds{1} - U_{\rm cc} + \mathds{1} \right. \nonumber \\
  &~~~~~~ \left. - U_{\rm cc}^\dagger - (\mathds{1} - U_{\rm cc}) (\mathds{1} - U_{\rm cc}^\dagger) \right] \nonumber \\
  &~~~~~~ \times (\mathds{1} - U_{\rm cc}^\dagger )^{-1} - \sum_{{\rm d}_i} \frac{1}{k_{\rm B}} \frac{\partial}{\partial T_{{\rm d}_i}} \langle \psi_{{\rm c}} \psi_{{\rm c}}^\dagger \rangle \nonumber \\
  &~~~ = 4 \left[ (\mathds{1} - U_{\rm cc}^\dagger )^{-1} + (\mathds{1} - U_{\rm cc} )^{-1} - \mathds{1} \right] \nonumber \\
  &~~~~~~ - \sum_{{\rm d}_i} \frac{1}{k_{\rm B}} \frac{\partial}{\partial T_{{\rm d}_i}} \langle \psi_{{\rm c}} \psi_{{\rm c}}^\dagger \rangle \mbox{.}
\end{align}
We used the unitarity of the scattering matrix, that is, Equation \eqref{eq:unitaryschem} on the second line and Equation \eqref{eq:aribtraryout} for the noise contribution to the output on the third line. 

We are interested in the off-diagonal part of the sensitivity matrix on the left-hand side (i.e., $(\partial/\partial T_{\rm s}) \langle \psi_{{\rm c}_i} \psi_{{\rm c}, j}^\ast \rangle$ with $i \neq j$), because this contains all the visibilities in a general interferometric setup. The first contribution on the right-hand side is nonzero only if $U_{\rm cc}$ is nondiagonal (i.e., there is a nonzero cross-talk between some readout channels). The second contribution is nonzero only if dissipative element(s) exist that can radiate into both relevant readout channels.

\section{Reciprocity for short dipoles}
\label{sec:reciprocity}

In this section we derive the relation between a short dipole antenna's transmission and reception properties. We assume that the dipole is that of Section \ref{sec:shortdipole} -- it has a capacitance $C$, conversion factor $\xi$, and is oriented along $\hat{\bm z}$. The factor $\xi$ converts between the charge on the dipole and its dipole moment. We will show that the same factor converts between an incident electric field and the voltage across the dipole in an open circuit. 

As a warmup, we consider an isolated capacitor with capacitance $C$ in a circuit with a battery of EMF $V$. If the capacitor acquires a charge $Q$, the stored energy is $E_{\rm c}(Q) = Q^2/(2 C)$. The work performed by the battery is $W_{\rm v}(Q) = V Q$. The system's {\em energy functional} represents the amount of energy dissipated in the process of charging and equals $\mathcal{E}(V, Q) = V Q - Q^2/(2C)$. The energy functional attains a local extremum when the stored charge and voltage obey the usual relation, $Q = C V$.

In our case, the capacitor develops an extra dipole moment, ${\bm p}(Q) = \xi Q \ \hat{\bm z}$. In an external electric field, $\bm E$, the stored energy is\footnote{We take the external electric field to be a boundary condition, and do not write down its energy density.} $E_{\rm c}(Q, \bm E) = Q^2/(2 C) - {\bm p}(Q) \cdot {\bm E} = Q^2/(2 C) - \xi Q E_{\rm z}$. The new energy functional, and the extremizing charge are
\begin{equation}
  \mathcal{E}(V, Q, {\bm E})  = V Q - \frac{Q^2}{2C} + \xi Q E_{\rm z} \mbox{,}
\end{equation}
implying
\begin{equation}
  \frac{\partial}{\partial Q}\mathcal{E}(V, Q, {\bm E})  = V - \frac{Q(V, {\bm E})}{C} + \xi E_{\rm z} = 0 \mbox{.} \label{eq:equivckt}
\end{equation}
This shows that the electric field's effect is equivalent to a voltage source in series with the capacitor, as shown in Figure \ref{fig:eqcircuit}.

\bibliography{references}

\end{document}